\documentclass[epss, 10pt]{article}
\input{psfig.sty}
\usepackage{amsmath}
\usepackage{xspace}
\usepackage{verbatim}
\usepackage{latexsym}
\usepackage{subfigure}
\usepackage{amssymb}
\usepackage{bm}
\usepackage{graphicx}
\usepackage{psfrag}
\usepackage{algorithm}
\usepackage{algorithmic}
\usepackage{authblk}
\begin{document}
\renewcommand{\textfraction}{0}

\title{Discrete denoising of heterogenous two-dimensional data  }
\author[1]{Taesup Moon}
\author[2]{Tsachy Weissman}
\author[2]{Jae-Young Kim}
\affil[1]{\small{Yahoo! Labs, 701 First Ave, Sunnyvale, CA 94089}}
\affil[2]{Stanford University, 350 Serra Mall, Stanford, CA 94305}
%\author{
%Taesup Moon\affiliation{Taesup Moon is with Yahoo! Labs, Sunnyvale, CA 94089}  \and Jae-Young Kim \footnote{Jae-Young Kim is with Stanford University, Stanford, CA 94305} \and Tsachy Weissman \footnote{} }
%\date
\maketitle \thispagestyle{empty}
\begin{abstract}
We consider discrete denoising of two-dimensional data with
characteristics that may be varying abruptly between regions.
 Using a quadtree decomposition technique and space-filling curves,
we extend the recently developed S-DUDE (Shifting Discrete Universal
DEnoiser), which was tailored to one-dimensional data, to the
two-dimensional case. Our scheme competes with a genie that has
access, in addition to the noisy data, also to the underlying
noiseless data, and can employ $m$ different two-dimensional sliding
window denoisers along  $m$ distinct regions obtained by a quadtree
decomposition with $m$ leaves, in a way that minimizes the overall
loss. We show that, regardless of what the underlying noiseless data
may be, the two-dimensional S-DUDE performs essentially as well as
this genie, provided that the number of distinct regions satisfies
$m=o(n)$, where $n$ is the total size of the data. The
resulting algorithm complexity is still linear in both $n$ and $m$, as in the one-dimensional case. Our experimental results show that the two-dimensional S-DUDE 
can be effective  when the characteristics of the underlying clean image vary across different regions in the data.\\

\end{abstract}

\normalsize

\newtheorem{claim}{Claim}
\newtheorem{guess}{Conjecture}
\newtheorem{defn}{Definition}
\newtheorem{fact}{Fact}
\newtheorem{assumption}{Assumption}
\newtheorem{theorem}{Theorem}
\newtheorem{lem}{Lemma}
\newtheorem{cor}{Corollary}
\newtheorem{proof}{Proof}
\newtheorem{pfth}{Proof of Theorem}
\newtheorem{ctheorem}{Corrected Theorem}
\newtheorem{corollary}{Corollary}
\newtheorem{proposition}{Proposition}
\newtheorem{example}{Example}
\newcommand{\mat}[2]{\ensuremath{
\left( \begin{array}{c} #1 \\ #2 \end{array} \right)}}
\newcommand{\ba}{\begin{align}
\end{align}}
\newcommand{\eq}[1]{(\ref{#1})}
\newcommand{\one}[1]{\ensuremath{\mathbf{1}_{#1}}}
\newcommand{\am}{\mbox{argmin}}
\newcommand{\dmin}{d_{\mbox{min}}}
\newcommand{\be}{\begin{eqnarray}}
\newcommand{\ee}{\end{eqnarray}}
\newcommand{\eps}{\varepsilon}
\newcommand{\imipi}{\int_{-\infty}^{\infty}}
\newcommand{\mug}{\stackrel{\triangle}{=}}
\renewcommand{\thesubsection}{\Alph{subsection}}
\def \bfpi  {\bm{\pi}}
\def \bflambda  {\bm{\lambda}}

\newcommand{\qed}{\quad\blacksqaure}
\newcommand{\Shat}{\hat{\mathbf{S}}}
\newcommand{\Ss}{\mathbf{S}^{*}}
\newcommand{\Sb}{\mathbf{S}}
\newcommand{\Skhat}{\hat{\mathbf{S}}_{k,m}}
\newcommand{\Sks}{\mathbf{S}_k^{*}}
\newcommand{\Skb}{\mathbf{S}_k}
\newcommand{\Snm}{\mathcal{S}_{0,m}^n}
\newcommand{\Sknm}{\mathcal{S}_{k,m}^n}
\newcommand{\tL}{\tilde{L}}
\newcommand{\mcA}{\mathcal{A}}
\newcommand{\mcB}{\mathcal{B}}
\newcommand{\mcS}{\mathcal{S}}
\newcommand{\mcX}{\mathcal{X}}
\newcommand{\Xb}{\mathbf{X}}
\newcommand{\mcXhat}{\hat{\mathcal{X}}}
\newcommand{\mcZ}{\mathcal{Z}}
\newcommand{\mcN}{\mathcal{N}}
\newcommand{\cb}{\mathbf{c}}
\newcommand{\Cb}{\mathbf{C}}
\newcommand{\integers}{\mathbb{Z}}
\newcommand{\naturals}{\mathbb{N}}
\newcommand{\Vmn}{V_{m\times n}}
\newcommand{\zmn}{z_{m\times n}}
\newcommand{\mtn}{m\times n}
\newcommand{\mtnn}{m{\times} n}  % same as \mtn but narrower
\newcommand{\mbPi}{\mathbf{\Pi}}
\newcommand{\mbm}{\mathbf{m}}
\newcommand{\mbz}{\mathbf{z}}
\newcommand{\mbu}{\mathbf{u}}
\newcommand{\xhmn}{\hat{x}_{\mtnn}}
\newcommand{\xmn}{x_{\mtnn}}
\newcommand{\Xmnuniv}{\hat{X}^{\mtnn}_{{\scriptsize{\sf univ}}}}
\newcommand{\mexp}[2]{#1{\cdot} 10^{#2}}
\renewcommand{\thesubsection}{\thesection.\arabic{subsection}}

\section{Introduction}

Discrete denoising is the problem of reconstructing the components
of a finite-alphabet sequence  based on the %\emph{entire}
observation of its Discrete Memoryless Channel (DMC)\footnote{The DMC is assumed  known throughout this paper. This assumption is benign in applications where the DMC is easily learnable from the data.}-corrupted
version.  %The quality of the reconstruction is evaluated via a
%user-specified (single-letter) loss function.
Universal discrete denoising, in which no statistical or
other properties are known a priori about the underlying clean data
and the goal is to attain optimum performance, was considered and
solved in \cite{Dude}. The main result in \cite{Dude} is the
\emph{semi-stochastic} setting one, which asserts that, regardless of
what the underlying individual sequence may be, the Discrete
Universal Denoiser (DUDE)  attains the performance of the
\emph{best} sliding window denoiser that would be chosen by a genie who
accesses,  in addition to the noisy sequence, the   underlying clean
data. Recently, in \cite{sdude}, a %meaningful
generalization %of \cite{Dude}
has been carried out for the case in which the characteristics of
the underlying sequence change  over time. The new scheme,
called Shifting Discrete Universal Denoiser (S-DUDE), was shown to
achieve the performance of the \emph{best combination} of sliding window
denoisers, allowing at most $m$ shifts (i.e., switches from one sliding window denoiser to another) along the sequence, provided
that $m$ grows sub-linearly in the data size $n$, regardless of what
the underlying noiseless sequence may be. It was also shown in
\cite{sdude} that the scheme can be implemented efficiently via dynamic programming, with
linear complexity both in $n$ and $m$.

One of the domains in which DUDE found its application is  image denoising. It was shown in the experimental results of \cite{Dude} and \cite{2d_dude} that DUDE achieves or often outperforms the best of several of the state-of-the-art image denoisers for small-alphabet images, many of which are sliding window schemes. It is natural then to attempt to extend S-DUDE for images, namely, two-dimensional data, as well.  The motivation is clear; images tend to have locally distinct characteristics, and allowing the sliding window denoisers to shift from one region to another may significantly improve the denoising performance compared to applying
one fixed sliding window denoiser throughout all the data. However, whereas the extension of the DUDE
to two-dimensional data was straightforward (cf.\ \cite[Section VIII-C]{Dude} and  \cite{2d_dude}), that of the S-DUDE is  highly non-trivial, since it requires  segmentation of the data, based on its noisy observation, into ׂhomogeneous regions׃ in a way that minimizes the
overall loss. Such segmentation is significantly more involved and often intractable, in contrast to the one-dimensional case of the S-DUDE, which only required to divide the data into distinct intervals with associated denoisers.

Due to this difficulty of general segmentation of data, we instead adopt a restricted, yet rich enough, segmentation scheme - quadtree decomposition - to build a reference class of shifting two-dimensional sliding window denoisers. Then, we employ the space-filling Peano-Hilbert curve \cite{LZ86,phcurves} to scan the data so that applying the original one-dimensional S-DUDE on the scanned data can achieve the best performance among the schemes in the reference class, regardless of the underlying clean data. The quadtree decomposition has been popular in image compression
\cite{quadtreeFeder}\cite{quadtreeCompression} and pattern
recognition \cite{quadtreePR}, and recently in \cite{quadtreeDenoising}, it has also
been applied to denoising continuous-valued signals by viewing denoising as a low-rate
lossy compression problem. The Peano-Hilbert curves have been used, among other applications, in universal compression of two-dimensional data in both the individual sequence setting \cite{LZ86} and the probabilistic setting \cite{WeissmanMannor}. A more general problem of scanning and predicting multi-dimensional data was considered in \cite{cohen1,cohen2}. The combination of the quadtree decomposition and the Peano-Hilbert scanning for discrete denoising problems is the main contribution of this paper. 

Our resulting denoising scheme, 2-D S-DUDE, still enjoys the performance guarantees that parallel those of \cite{sdude} for two-dimensional data. That is, regardless of what the underlying clean data might be, 2-D S-DUDE performs asymptotically as well as the best combination of the two-dimensional sliding window denoisers that can shift across at most $m$ distinct regions, as can be segmented by the best quadtree decomposition. Our use of the Peano-Hilbert scan is essential to obtain a scheme of which complexity remains linear in both the data size  and the number of distinct segments $m$,  whereas an effort of directly finding the best quadtree decomposition may have resulted in a scheme with complexity exponential in $m$. We show the effectiveness of our scheme by experimental results that demonstrate 2-D S-DUDE outperforming 2-D DUDE, particularly for images of space-varying characteristics, such as scanned magazines, etc.

The rest of the paper is organized as follows: Section \ref{sec
notation}  collects necessary notation, preliminary results and
detailed motivation for this work. Our main theoretical results and algorithm are given in Section
\ref{section: main result}, and the experimental results are presented in Section \ref{sec: experiments}. Concluding remarks with a discussion
of future work are given in Section \ref{sec: conclusion}.

\section{Notation, Preliminaries, and Motivation}\label{sec notation}
\subsection{Notation}\label{subsec: notation}

We follow the notation of \cite{sdude}. Let $\mcX,\mcZ,\mcXhat$
denote, respectively, the alphabet of the clean, noisy, and
reconstructed sources, which are assumed to be finite. As in
\cite{Dude, sdude, UFP06}, the noisy sequence is a
DMC-corrupted version of the clean one, where the channel matrix is denoted by
$\mathbf{\Pi}=\{\Pi(x,z)\}_{x\in\mcX,z\in\mcZ}$, and $\Pi(x,z)$ stands for 
the probability of a noisy symbol $z$ when the underlying clean
symbol is $x$. Throughout the paper, $\mathbf{\Pi}$ is assumed to be known and fixed, and of full row rank. 
%The $z$-th column of
%$\mathbf{\Pi}$ will be denoted as $\pi_z$. 
When a reconstruction
$\hat{x}$ is made for a clean symbol $x$, the goodness of the
reconstruction is measured by a loss function
$\Lambda:\mcX\times\mcXhat\rightarrow[0,\infty)$. Upper case letters
 denote random variables; lower case letters 
denote either individual deterministic quantities or specific
realizations of random variables.
%For simplicity,
%from now on, we will assume that the alphabet size of
%$\mcX,\mcZ,\mcXhat$ is all equal to $M$ and $\mathbf{\Pi}$ is an
%$M$-by-$M$ invertible matrix.
Without loss of generality, the elements of any finite alphabet
$\mathcal{V}$ will be identified with
$\{0,1,\cdots,|\mathcal{V}|-1\}$. For $\mathcal{V}$-valued sequence,
we let $v^n=(v_1,\cdots,v_n)$, $v_m^n=(v_m,\cdots,v_n)$, and
$v^{n\backslash t}=v^{t-1}v_{t+1}^n$.
%
%
%
%We let $\mathcal{V}^{\infty}$ denote the set of one-sided infinite
%sequences with $\mathcal{V}$-valued components, i.e.,
%$\mathbf{v}\in\mathcal{V}^\infty$ is of the form
%$\mathbf{v}=(v_1,v_2,\cdots),v_i\in\mathcal{V},i\geq 1$. For
%$\mathbf{v}\in\mathcal{V}^{\infty}$, let $v^n=(v_1,\cdots,v_n)$ and
%$v_m^n=(v_m,\cdots,v_n)$. Furthermore, we let $v^{n\backslash t}$
%denote the sequence $v^{t-1}v_{t+1}^n$.
$\mathbb{R}^{\mathcal{V}}$ is a space of $|\mathcal{V}|$-dimensional
column vectors with real-valued components indexed by the elements
of $\mathcal{V}$. 
%The $a$-th component of
%$q \in \mathbb{R}^{\mathcal{V}}$ will be denoted %either by $v[a]$ or
%by either $q_a$ or $q[a]$.
%%(according to what will result in an overall simpler
%%expression in each particular case.)
%When $\Gamma$ is a $|\mathcal{V}_1|\times|\mathcal{V}_2|$ matrix,
%$\Gamma_{\max}$ stands for
%$\max_{v_1\in\mathcal{V}_1,v_2\in\mathcal{V}_2}\Gamma(v_1,v_2)-\min_{v_1\in\mathcal{V}_1,v_2\in\mathcal{V}_2}\Gamma(v_1,v_2)$.
%In addition, $\mathbf{1}_{\{\cdot\}}$ denotes an indicator of the
%event inside $\{\cdot\}$.

%The goodness of the reconstruction is measured by a given loss
%function $\Lambda:\mcX\times\mcXhat\rightarrow[0,\infty)$, where the
%maximum single-letter loss is denoted by $\Lambda_{\max}$, and
%$\lambda_{\hat{x}}$ denotes the $\hat{x}$-th column of the loss
%matrix.

Now, consider the set $\mcS=\{s:\mcZ\rightarrow\mcXhat\}$, which is
the (finite) set of mappings that take $\mcZ$ into $\mcXhat$. We
refer to elements of  $\mcS$ as ``single-symbol denoisers'', since
each $s \in \mcS$ can be thought of as a rule for estimating $x \in
\mathcal{X}$ on the basis of $z \in \mathcal{Z}$. Then, for any
$s\in\mcS$, we can always devise an estimated loss $\ell(Z,s)$ with the knowledge of $\mathbf{\Pi}$, which is an unbiased estimate of the true expected loss $E_x\Lambda(x,s(Z))$, i.e.\ satisfying  
\begin{eqnarray}
E_x\ell(Z,s) = E_x\Lambda(x,s(Z)) \ \ \ \forall x \in
\mathcal{X}. \label{estimated loss}
\end{eqnarray}
The expectation in (\ref{estimated loss}) is with respect to the conditional distribution on $Z$ given $x$, $\Pi(x,\cdot)$. 
For more details on the motivation for using a loss function $\ell$ satisfying  (\ref{estimated loss}), and on its explicit form, readers may refer to \cite[Section II-A]{sdude}.

%
%an unbiased estimator for $\Lambda(x,s(Z))$ (based on
%$Z$ only), where $x$ is a deterministic symbol and $Z$ is the output
%of the DMC when the input is $x$, can be obtained as in
%\cite{UFP06}. First, pick a function
%$h:\mcZ\rightarrow\mathbb{R}^{\mcX}$ with the property that, for
%$a,b\in\mcX$,
%\begin{align}
%E_ah_b(Z)=&\sum_{z\in\mcZ}h_b(z)\Pi(a,z)=\mathbf{1}_{\{a=b\}},\nonumber
%%         =&\delta(a,b)\triangleq \left\{
%%\begin{array}{cc}
%%1, & \textrm{if $a=b$} \\
%%0, & \textrm{otherwise}\\
%%\end{array}\right\},\label{h def}
%\end{align}
%where $E_a$ denotes expectation over the channel output $Z$  when
%the underlying channel input is $a$, and $h_b(z)$ denotes the $b$-th
%component of $h(z)$. As in \cite{UFP06}, the guarantee of the
%existence of such $h$ follows from the full row rank assumption of
%the channel matrix. Now, for any $s\in\mcS$,
%$\rho(s)\in\mathbb{R}^{\mcX}$ denotes the column vector with $x$-th
%component $$
%\rho_x(s)=\sum_{z}\Lambda(x,s(z))\Pi(x,z)=E_x\Lambda(x,s(Z)) . $$ In
%words, $\rho_x(s)$ is the expected loss using the single-symbol
%denoiser $s$, while the underlying symbol is $x$.
% Considering $\mcS$ as an action
%space alphabet, we define a loss function
%$\ell:\mcZ\times\mcS\rightarrow\mathbb{R}$ as \be
%\ell(z,s)=h(z)^T\cdot\rho(s). \label{est loss} \ee We observe from
%the definition of $h$ and $\rho$ that $\ell(Z,s)$ is an unbiased
%estimate of $\Lambda(x,s(Z))$ since $ E_x\ell(Z,s)= E_x
%h(Z)^T\cdot\rho(s) =  E_x\Lambda(x,s(Z))$ for all $x\in\mcX. $

%

\subsection{DUDE and S-DUDE for 1-D data}\label{prelim}

Here, we review and summarize the results from
\cite{Dude} and \cite{sdude}, and collect the ideas that will be
needed for this paper. For one dimensional data, an \emph{$n$-block
denoiser} is a collection of $n$ mappings
$\hat{\mathbf{X}}^n=\{\hat{X}_t\}_{1\leq t\leq n}$, where
$\hat{X}_t:\mcZ^n\rightarrow \mcXhat$. The performance of the
denoiser $\hat{\mathbf{X}}^n$ on the individual sequence pair
$(x^n,z^n) $ is measured by its normalized cumulative loss $$
L_{\hat{\mathbf{X}}^n}(x^n,z^n)=\frac{1}{n}\sum_{t=1}^n\Lambda(x_t,\hat{X}_t(z^n)).
$$
%
%normalized cumulative loss of the denoiser $\hat{\mathbf{X}}^n$ on
%the individual sequence pair $(x^n,z^n) $ is represented as $
%L_{\hat{\mathbf{X}}^n}(x^n,z^n)=\frac{1}{n}\sum_{t=1}^n\Lambda(x_t,\hat{X}_t(z^n)).
%$
%In words, $L_{\hat{\mathbf{X}}^n}(x^n,z^n)$ is the normalized
%(per-symbol) loss, as measured under the loss function $\Lambda$,
%when using the denoiser $\hat{\mathbf{X}}^n$ and when the observed
%noisy sequence is $z^n$ while the underlying clean one is $x^n$. %The
%notation $L_{\hat{\mathbf{X}}^n}$ is extended for $1\leq i\leq j\leq
%n$, $
%L_{\hat{\mathbf{X}}^n}(x_i^j,z^n)=\frac{1}{j-i+1}\sum_{t=i}^j\Lambda(x_t,\hat{X}_t(z^n)),
%$ denoting the normalized (per-symbol) loss between (and including)
%locations $i$ and $j$.
As argued in \cite[Section II-B]{sdude}, the $n$-block denoiser
$\hat{\mathbf{X}}^n=\{\hat{X}_t\}_{1\leq t \leq n}$ can be
identified with $\mathbf{F}^n=\{F_t\}_{1\leq t\leq n}$, where
$F_t:\mcZ^{n\backslash t}\rightarrow \mcS$ is defined as follows:
$F_t(z^{n\backslash t},\cdot)$ is the single-symbol denoiser in
$\mcS$ satisfying
\begin{align}
\hat{X}_t(z^n)=F_t(z^{n\backslash t},z_t), \ \ \ \ \forall
z_t.\label{F map}
\end{align}
%Therefore, we can adopt the view that at each time $t$, an $n$-block
%denoiser is choosing a single-symbol denoiser based on all the noisy
%sequence components but $z_t$, and applying that single-symbol
%denoiser on $z_t$ to yield the $t$-th reconstruction $\hat{x}_t$.
%Conversely, any sequence of mappings into single-symbol denoisers
%$\mathbf{F}^n$ defines a denoiser $\hat{\mathbf{X}}^n$, again via
%\eq{F map}. This viewpoint will be useful in what follows.
One special class of widely used $n$-block denoisers is that of the
$k$-th order ``sliding window'' denoisers, which we denote by
$\hat{\mathbf{X}}^{n,\mcS_k}$. Such denoisers are of the form \be
\hat{X}_t^{s_k}(z^n)=s_k(z_{t-k}^{t+k})=s_k(\cb_t,z_t)\label{def:
sliding window denoiser} \ee for $t=k+1,\cdots,n-k$, where $s_k$ is
an element of $\mcS_k=\{s_k:\mcZ^{2k+1}\rightarrow\mcXhat\}$, the
(finite) set of mappings from $\mcZ^{2k+1}$ into $\mcXhat$
%\footnote{The value of $\hat{X}_t^{s_k}(z^n)$ for $t\leq k$ and
%$t>n-k$ is defined, for concreteness and simplicity, as an arbitrary
%fixed symbol in $\hat{\mathcal{X}}$.}
, and $\cb_t\triangleq(z_{t-k}^{t-1},z_{t+1}^{t+k})$ is the
(two-sided) $k$-th order context for $z_t$. We refer to
$s_k\in\mcS_k$ as a ``$k$-th order denoiser''. Note that
$\mcS_0$ equals to $\mcS$ in the previous section. 
$\mathbf{C}_k\triangleq\{(u_{-k}^{-1},u_{1}^{k}):(u_{-k}^{-1},u_{1}^{k})\in\mcZ^{2k}\}$
is  the set of all possible $k$-th order contexts, and for given $z^n$
and each $\cb\in\Cb_k$, $$
\mathcal{T}(\mathbf{c})\triangleq\big\{\tau:\mathbf{c}_{\tau}=\mathbf{c},
\quad k+1\leq \tau\leq n-k\big\}$$ is further defined to be the set of indices 
where the context of $z_i$ equals $\mathbf{c}$. From the association in
(\ref{F map}) and the definition (\ref{def: sliding window
denoiser}), we can deduce that for each $\cb\in\Cb_k$, the $k$-th
order sliding window denoiser employs a time-invariant single-symbol
denoiser, $s_k(\cb,\cdot)$, at all points $t\in\mathcal{T}(\cb)$.
%In other
%words, the $k$-th order sliding window denoiser partitions $z^n$
%into the subsequences associated with the various contexts, and
%employs a time-invariant single-symbol scheme on each such
%subsequence.

In \cite{Dude}, the performance target of the denoising is $$
D_k(x^n,z^n)\triangleq
%\min_{\hat{\mathbf{X}}^n\in\hat{\mathbf{X}}^{n,\mcS_k}}L_{\hat{\mathbf{X}}^n}(x_{k+1}^{n-k},z^n)\nonumber\\
%            =&
            \min_{s_k\in\mcS_k}\frac{1}{n-2k}\sum_{t=k+1}^{n-k}\Lambda(x_t,s_k(\cb_t,z_t)),%\label{Dmin}
           % &=&\frac{1}{n-2k}\sum_{(u_{-k}^{-1},u_1^k)\in\mcZ^{2k}}\bigg[\min_{s\in\mcS}\sum_{t\in\mathcal{T}(u_{-k}^{-1},u_1^k)}\Lambda(x_t,s(z_t))\bigg]
$$ the minimum normalized loss on $(x^n,z^n)$ that can be attained by
a $k$-th order sliding window denoiser.
%Since (\ref{Dmin}) can be
%expressed as
%$\frac{1}{n-2k}\sum_{\cb\in\Cb_k}\big[\min_{s\in\mathcal{S}}\sum_{\tau\in\mathcal{T}(\cb)}\Lambda(x_{\tau},s(z_{\tau}))\big]$,
We can easily verify that for each $\cb\in\Cb_k$, the best $k$-th
order sliding window denoiser that achieves $D_k(x^n,z^n)$ will
employ the single-symbol denoiser \be
\arg\min_{s\in\mcS}\sum_{\tau\in\mathcal{T}(\cb)}\Lambda(x_\tau,s(z_\tau)),\label{argmin
true} \ee at all points $t\in\mathcal{T}(\cb)$, which is determined
from the joint empirical distribution of pairs
$\{(x_\tau,z_\tau):\tau\in\mathcal{T}(\cb)\}$. It is shown in
\cite[Theorem 1]{Dude} that, despite the lack of knowledge of $x^n$,
$D_k(x^n,Z^n)$ is essentially achievable by the Discrete Universal
DEnoiser (DUDE), which accesses only $Z^n$ and is implementable with
linear complexity in $n$.
%
%
%a linear complexity algorithm, the Discrete Universal DEnoiser
%(DUDE), that only has
%access to $Z^n$. %
%
%
%
%This scheme has a simple linear complexity in $n$, and is dubbed as
%the Discrete Universal DEnoiser (DUDE),
%$\hat{\mathbf{X}}^{n,k}_{\mathrm{univ}}$.
For each $\cb\in\Cb_k$, the DUDE can be shown to employ \be
\arg\min_{s\in\mcS}\sum_{\tau\in\mathcal{T}(\cb)}\ell(z_\tau,s)\label{argmin
est}\ee at all points $t\in\mathcal{T}(\cb)$, where $\ell(z,s)$ is
the loss function chosen to satisfy \eq{estimated loss}. By comparing \eq{argmin
true} with \eq{argmin est}, we observe that, for each context, the DUDE merely works
with the estimated loss $\ell(z_{\tau},s)$ in lieu of the
true loss $\Lambda(x_{\tau},s(z_{\tau}))$.
%Following theorem was
%proved to show that working with $\ell(Z_{\tau},s)$ is enough for
%essentially achieving the genie-aided performance $D_k(x^n,Z^n)$.

%\begin{theorem}(\cite[Theorem 1]{Dude})\label{dude main thm}
%Take $k=k_n$ satisfying $k_n|\mcZ|^{2k_n}=o(n/\log n)$. Then, for
%all $\mathbf{x}\in\mcX^{\infty}$, the sequence of denoisers
%$\{\hat{\mathbf{X}}^{n,k_n}_{\mathrm{univ}}\}$ satisfies:
%\begin{enumerate}
%\item[a)]
%$\lim_{n\rightarrow\infty}\Big[L_{\hat{\mathbf{X}}^{n,k_n}_{\mathrm{univ}}}(x^n,Z^n)-D_{k_n}(x^n,Z^n)\Big]=0\quad\textrm{a.s.}$
%%provided that $k_n|\mcZ|^{2k_n}=o(n/\log n)$.
%\item[b)]
%$
%E[L_{\hat{\mathbf{X}}^{n,k_n}_{\mathrm{univ}}}(x^n,Z^n)-D_{k_n}(x^n,Z^n)]=O\big(\sqrt{\frac{k_n|\mcZ|^{2k_n}}{n}}\big).
%$
%\end{enumerate}\vspace{.05in}
%\end{theorem}

The idea of working with estimated loss to achieve the genie-aided
performance  has been adopted again in \cite{sdude} to refine the
result of \cite{Dude}. The main motivation of \cite{sdude} was the
observation that when the characteristics of the underlying sequence
$x^n$ change with time, allowing the $k$-th order denoiser to 
change from one interval of the data to another can
further reduce the overall loss significantly. %
%it is reasonable that the $k$-th order denoisers can also be
%time-varying throughout the noisy sequence $z^n$ to further minimize
%the overall normalize loss.
Therefore, whereas the DUDE competed with the best \emph{fixed}
$k$-th order denoiser, \cite{sdude} competes with the best among
$\mathcal{S}_{k,m}^n$, a set of ``combinations'' of $k$-th order
denoisers $\{s_{k,t}\}_{t=k+1}^{n-k}$ that allow at most $m$ shifts
within $t\in\mathcal{T}(\cb)$ for each $\cb\in\Cb_k$.
%In \cite{sdude}, this result was extended to the case where the
%time-varying nature of the underlying sequence $x^n$ was taken into
%account. As shown above, the $k$-th order sliding window denoiser
%applies a time-invariant single-symbol denoiser for the time indices
%where the $k$-th order contexts are identical. However, when the
%characteristics of the underlying sequence $x^n$ changes abruptly
%with time, allowing the $k$-th order denoisers to shift over time
%would further minimize the total loss. In \cite{sdude},
%$\mathcal{S}_{k,m}^n$ was defined to represent a set of combinations
%of $k$-th order denoisers $\{s_{k,t}\}$'s that allow at most $m$
%shifts along the subsequences that share the same contexts.
Thus, \cite{sdude} sets a more ambitious performance target 
\begin{eqnarray}
D_{k,m}(x^n,z^n)\triangleq\min_{\mathbf{S}\in\mathcal{S}_{k,m}^n}\frac{1}{n-2k}\sum_{t=k+1}^{n-k}\Lambda(x_t,s_{k,t}(\cb_t,z_t)),\label{sdude performance target}
\end{eqnarray}
 the minimum normalized loss on $(x^n,z^n)$ that can be achieved by
the sequence of $k$-th order denoisers that allow at most $m$ shifts (changes)
within each context. It is clear that $D_{k,m}(x^n,z^n)\leq
D_k(x^n,z^n)$ for all $(x^n,z^n)$. The new algorithm devised in
\cite{sdude} was called the $(k,m)$-Shifting Discrete Universal
DEnoiser (S-DUDE), $\hat{\mathbf{X}}_{\textrm{univ}}^{n,k,m}$, and
was able to asymptotically achieve $D_{k,m}(x^n,Z^n)$ on the basis
of $Z^n$ only, provided that $m$ grows sub-linearly in $n$. The key
trick was again to work with the estimated loss $\ell(z,s)$ to
obtain
\begin{eqnarray}
\hat{\mathbf{S}}_{k,m}=\arg\min_{\mathbf{S}\in\mathcal{S}_{k,m}^n}\frac{1}{n-2k}\sum_{t=k+1}^{n-k}\ell(z_t,s_{k,t}(\cb_t,\cdot)),\label{sdude
def}
\end{eqnarray}
and employ it throughout the sequence. The following theorem shows
that by utilizing the estimated loss, we can successfully learn the
best (at most $m$) shifts of $k$-th order denoisers throughout $z^n$
to minimize the overall normalized loss.
\begin{theorem}(\cite[Theorem 4]{sdude})\label{sdude main}
Suppose $k=k_n$ and $m=m_n$ grow with $n$ sufficiently slowly to satisfy conditions detailed in
\cite[Claim 1]{sdude}. Then, for all
$\mathbf{x}\in\mathcal{X}^{\infty}$, the sequence of denoisers
$\{\hat{\Xb}^{n,k,m}_{\textrm{univ}}\}$ satisfies
\begin{enumerate}
\item[a)]\label{sdude a}
$\lim_{n\rightarrow\infty}\Big[L_{\hat{\mathbf{X}}^{n,k,m}_{\mathrm{univ}}}(x^n,Z^n)-D_{k,m}(x^n,Z^n)\Big]=0\quad\textrm{a.s.}$

\item[b)] For any $\delta>0$,
$
E[L_{\hat{\mathbf{X}}^{n,k,m}_{\mathrm{univ}}}(x^n,Z^n)-D_{k,m}(x^n,Z^n)]=O\big(\sqrt{k_n|\mcZ|^{2k_n}(\frac{m_n}{n})^{1-\delta}}\big).
$
\end{enumerate}\vspace{.0in}
\end{theorem}
Besides the performance guarantees, another key component of \cite{sdude} was developing an algorithm that can implement (\ref{sdude def}) efficiently, i.e., with complexity linear in both $n$ and $m$.  
%In parallel to the DUDE, implementing (\ref{sdude def}) can also be
%done in an efficient way, which only requires linear complexity in
%both $n$ and $m$. 
The details can be found in \cite[Section
V-A]{sdude}.

\subsection{Motivation}\label{subsec: motivation}

%The S-DUDE turns out to treat each subsequence associated with each
%context separately,  finding the best combination of single-symbol
%denoisers with $m$ shifts within each subsequence, which generalizes
%what the DUDE did in (\ref{argmin est}).
%
%
%Just as the DUDE treats each subsequence associated with each
%context separately and finds the best single-symbol denoiser as in
%(\ref{argmin est}), it turns out that the S-DUDE also separately
%employs the same algorithm to find the best combination of
%single-symbol denoisers, on each subsequence associated with each
%context.
As pointed out in \cite[Section V-B]{sdude}, both DUDE and S-DUDE run the same algorithm in parallel along each subsequence associated with each context, and this characteristic enables us to extend both schemes to the two-dimensional (2-D) data case: use 2-D contexts and again run the algorithms on each subsequence. 
%this characteristic of
%running the same algorithm in parallel along each subsequence
%enables us to extend both schemes to the case of two-dimensional
%(2-D) data: run the same algorithm along each subset of data points
%associated with each two-dimensional context. 
However, as noted in
the Introduction, the extension to  2-D data of the S-DUDE is not as
straightforward as that of the DUDE. The main reason is that, whereas the output of DUDE is 
independent of the ordering of data within each context and only requires the empirical distribution of the data, the ordering of said data is very consequential for S-DUDE's output and its performance. 

\begin{figure}[h]
\centering
\includegraphics[width=0.25\textwidth]{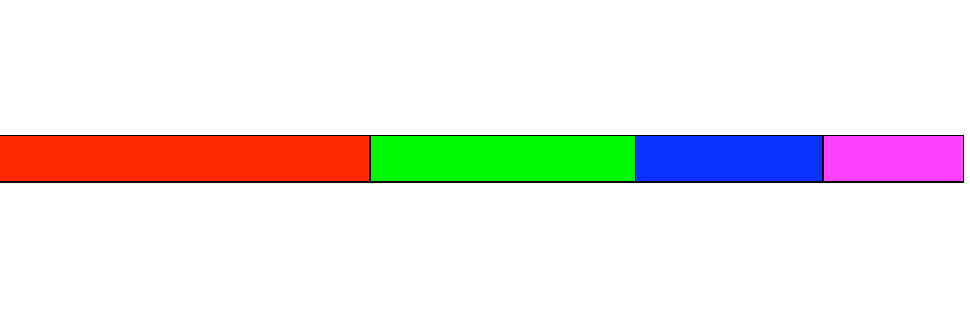}
\caption{Segmentation of 1-D data}\label{1d seg}
\end{figure}

\begin{figure}[h]
\centering
\includegraphics[width=0.2\textwidth]{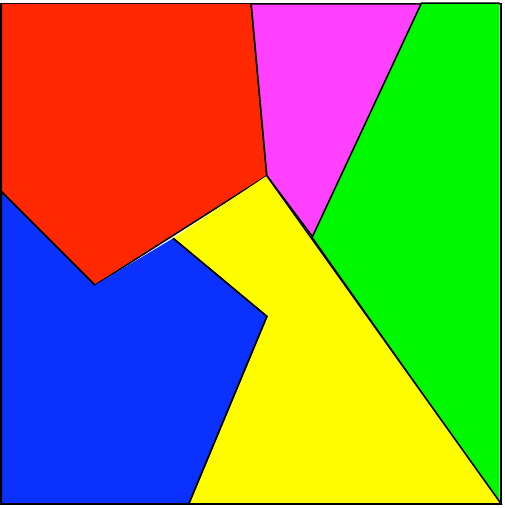}
\caption{Segmentation of 2-D data}\label{2d seg}
\end{figure}

The ordering of data is naturally given and fixed in one-dimensional (1-D) data. Therefore, S-DUDE only had to find shifting points based on noisy data so that applying different sliding window denoisers in different intervals will minimize the overall loss. Figure \ref{1d seg} shows one such segmentation in which different colors represent intervals where different sliding window denoisers are applied. 
In contrast, in the 2-D case, it is not clear how the 2-D version of S-DUDE should  segment into \emph{homogeneous} 2-D regions, instead of intervals, in order to allow shifting of sliding window denoisers across the data. As depcited in Figure \ref{2d seg}, the optimal segmentation that leads to the minimum loss can be  arbitrary, and hence, trying to learn the best segmentation solely based on \emph{noisy} data  would be overly ambitious.
One naive approach to avoid such a 2-D segmentation issue would be to first raster scan the image, then apply the ordinary S-DUDE on the resulting 1-D data, which was the method used in \cite[Section VI-A]{sdude}. However, this could often result in poor performance of the scheme since it may require the S-DUDE to shift too frequently, i.e., $m$ to become linear in the data size $n$,  which violates the necessary condition specified in \cite[Theorem 5]{sdude} for the scheme to work. This point can be seen by imagining the situation of running the raster scan vertically on the image in \cite[Figure 2]{sdude}, where even though the image consists of a small number of two-dimensional regions, when raster scanned into one-dimensional data the number of changes of the data characteristics grows significantly. 

To address this issue   of segmenting and scanning of the 2-D data in general, in this paper we focus on a more regularized class of segmentation schemes, the quadtree decomposition, to build the reference class of shifting sliding window denoisers rich enough for the denoising task at hand. Then, in order to compete with the reference class, we utilize the space-filling Peano-Hilbert (PH) curve to scan subsequence points for each 2-D context and run the ordinary S-DUDE on the P-H scanned, 1-D data. In the next section, we describe our scheme formally, and prove a performance guarantee for it, which parallels that of the scheme for 1D-data in \cite{sdude}.

\section{S-DUDE for 2-D data }\label{section: main
result}

Before presenting the 2-D extension of S-DUDE, we introduce additional notation in Section \ref{subsection: two dimensional context} through Section \ref{subsection: ph curve}. Then, in Section \ref{subsection: main results}, we derive our scheme and present theoretical guarantees of its performance. In Section \ref{subsection: algorithm}, we succinctly describe the algorithm and its complexity.

%
%We now formally present the 2-D extension of S-DUDE. Section \ref{subsection: two dimensional context} through Section \ref{subsection: ph curve}, we introduce some additional notations that are necessary for describing our scheme. Then, in Section \ref{subsection: main results}, we show theoretical results on the performance guarantee of the scheme, followed by the algorithmic summary in Section \ref{subsection: algorithm}.

% We restrict our attention to the reference class of quadtree (QT)
%decomposed denoisers, which segments the data into $m$ disjoint
%quadrants and applies a fixed denoiser in each quadrant. Limited as
%this class of denoisers may be,  the best combination among this
%class can significantly improve on the performance of the best fixed
%sliding window denoiser (which the 2-D DUDE is competing with),
%particularly for data with characteristics that differ between
%quadrants. We introduce some additional notation in subsection
%\ref{subsection: two dimensional context} and \ref{subsection:
%quadtree}, and present our main result in subsection
%\ref{subsection: main results}. Subsection \ref{subsection:
%algorithm} discusses the algorithm and its complexity.

\subsection{2-D data and contexts}\label{subsection: two
dimensional context}

We represent the 2-D data with the coordinate of each data point. For simplicity, we assume the 2-D data is always in the square form\footnote{For other cases, we can simply fill in remaining regions with dummy symbols.}. Then, denote $\mathcal{T}_N\triangleq\{t\in\mathbb{Z}^2:t=(t_1,t_2), 1\leq
t_1\leq N,1\leq t_2\leq N\} $ as the set of coordinates of the given
2-D data. Also, let $n=|\mathcal{T}_N|=N\times N$ be the total size of
the data. For $t\in\mathcal{T}_N$, $z_t$ will
denote the noisy symbol at location $t=(t_1,t_2)$, and $x^{N\times N}$ and $z^{N\times N}$ will denote the total clean and noisy 2-D data, respectively.   
In
addition, for $t\in\mathcal{T}_N$, $z^{N\times N\backslash t}$ will denote
$\{z_i: i\in\mathcal{T}_N, i\neq t\}$. With this notation, notions of the 2-D $n$-block denoiser
$\hat{\mathbf{X}}^{n}_{2D}$, the normalized cumulative loss
$$
L_{\hat{\mathbf{X}}^n_{2D}}(x^{N\times N},z^{N\times N})=\frac{1}{n}\sum_{t\in\mathcal{T}_N}\Lambda(x_t,\hat{X}_{t,\text{2D}}(z^{N\times N})),
$$
and the association in (\ref{F
map})  follow naturally in parallel to the 1-D data
case.

%We will overload the notation of the one-dimensional case given in Section \ref{subsec: notation}
%to the 2-D setting. 
%

%
%For example, for $t\in\mathcal{T}_N$, $z_t$ will
%denote the noisy symbol at location $(t_1,t_2)$, and $z^n$ will
%denote $\{z_t:t\in\mathcal{T}_N\}$, the total noisy data. In
%addition, for $t\in\mathcal{T}_N$, $z^{n\backslash t}$ will denote
%$\{z_i: i\in\mathcal{T}_N, i\neq t\}$. With this slightly overloaded
%notation, notions of the 2-D $n$-block denoiser
%$\hat{\mathbf{X}}^n$, the normalized cumulative loss
%$L_{\hat{\mathbf{X}}^n}(x^n,z^n)$, and the association of (\ref{F
%map}) would follow naturally in parallel to the one-dimensional data
%case.

\begin{figure}[h]
\begin{center}
\includegraphics[width=0.25\textwidth]{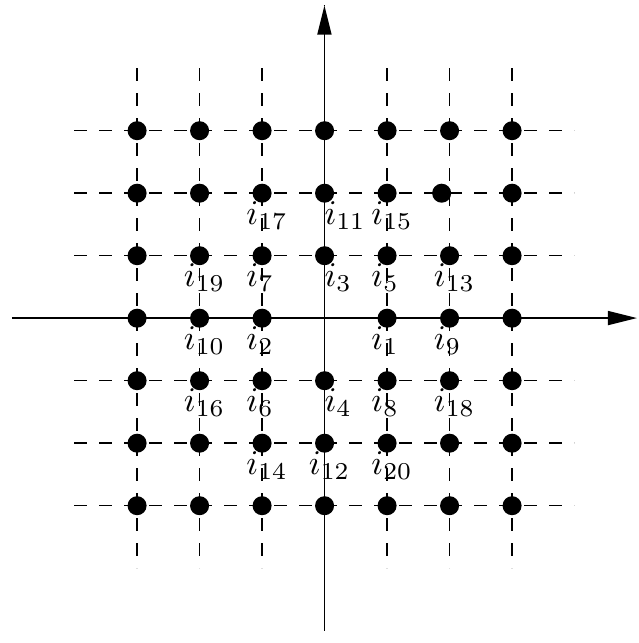}
\end{center}
\vspace{-.1in} \caption{The ordering coordinates of the 2-D integer
lattice plane}\label{2d contexts order}
\end{figure}

The 2-D $k$-th order sliding window denoisers can be understood
similarly. First, consider the sequence of coordinates,
$\mathcal{I}=(i_1,i_2,i_3,\cdots)$, in the 2-D lattice of integers,
in which the coordinates are enumerated in the order of increasing
distance to the origin as in Figure \ref{2d contexts order}.
Then, 2-D $k$-th order context for $z_t$ is defined to be
$\cb_{t,2D}=(z_{t+i_1},\cdots,z_{t+i_{2k}})$, where $(i_1,\cdots,i_{2k})$
are the first $2k$ coordinates of $\mathcal{I}$, and the additions of coordinates simply boil down to the translation of the coordinates. We also denote $\mathbf{C}_{k,\text{2D}}$ as a set of all possible 2-D $k$-th order contexts. Then, 2-D $k$-th
order sliding window denoiser at location $t$ is again of the form
$$\hat{X}_{t,\text{2D}}^{s_k}(z^n)=s_k(\cb_{t,\text{2D}},z_t),$$ 
with $\cb_{t,\text{2D}}\in\Cb_{k,\text{2D}}$, parallelling  \eq{def:
sliding window denoiser}.

%
%We denote $z_{t}$ as the noisy symbol at location $(t_1,t_2)$ of the
%data.  We will overload the notation $z^n$ and $z^{n\backslash t}$
%to denote the
%
%$\mathcal{I}=i_1i_2i_3\cdots$
%
%$\cb_t=(z_{t+i_1},\cdots,z_{t+i_{2k}})$
%
%%$\tb^k=(\tb_1,\tb_2,\cdots,\tb_k)$
%%
%%$t+\tb^k=(t+\tb_1,t+\tb_2,\cdots,t+\tb_k)$
%%
%%$z_{\tb+\tb^k}$
%
%
%$n=|I_N|=N\times N$.
%
%$z^{\mathcal{T_N}}$ denotes the noisy image.
%
%$z^{\mathcal{T_N}\backslash i}=\{z_\ell:\ell\in I_N, \ell\neq i\}$
%
%The $n$-th order denoiser is denoted as
%$$
%\hat{X}_i(z^n):\mcZ^{n}\rightarrow\mcXhat
%$$
%
%$\mathcal{O}$

\subsection{The quadtree (QT) decomposition}\label{subsection:
quadtree} 

A quadtree (QT) is a tree of which every node is either a
leaf or a parent node with four children. This QT structure can be
used to segment the 2-D data as follows. Each node of a QT at depth
$d$ represents a quadrant of size $\frac{N}{2^d}\times\frac{N}{2^d}$
(assuming $N=2^r$ and $d\leq r$), and a child node represents one of
the four quadrants inside the parent node's quadrant. The four
children of a parent node associate with the four quadrants of the
parent node's quadrant in the order of (upper-left
quadrant)$\rightarrow$(upper-right
quadrant)$\rightarrow$(lower-right quadrant)$\rightarrow$(lower-left
quadrant). Obviously, the root node is associated with the whole 2-D
data. The leaves of a QT represent the final segmentation of the
2-D data for given QT. Thus, if a QT has $m$ leaves, the resulting
segmentation has $m$ distinct regions.  For example, in Figure
\ref{image decomposition}, the two dimensional data is segmented into $13$
different regions, and the corresponding QT in Figure \ref{quadtree decomposition} has $m=13$ leaves.
\begin{figure}[h]
\centering \subfigure[2-D data decomposition]{\label{image
decomposition}
\includegraphics[width=0.2\textwidth]{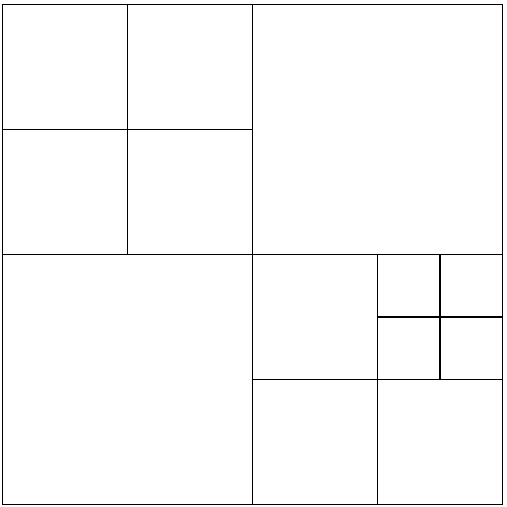}}
\hspace{.2in} \subfigure[QT decomposition]{\label{quadtree
decomposition}
\includegraphics[width=0.2\textwidth]{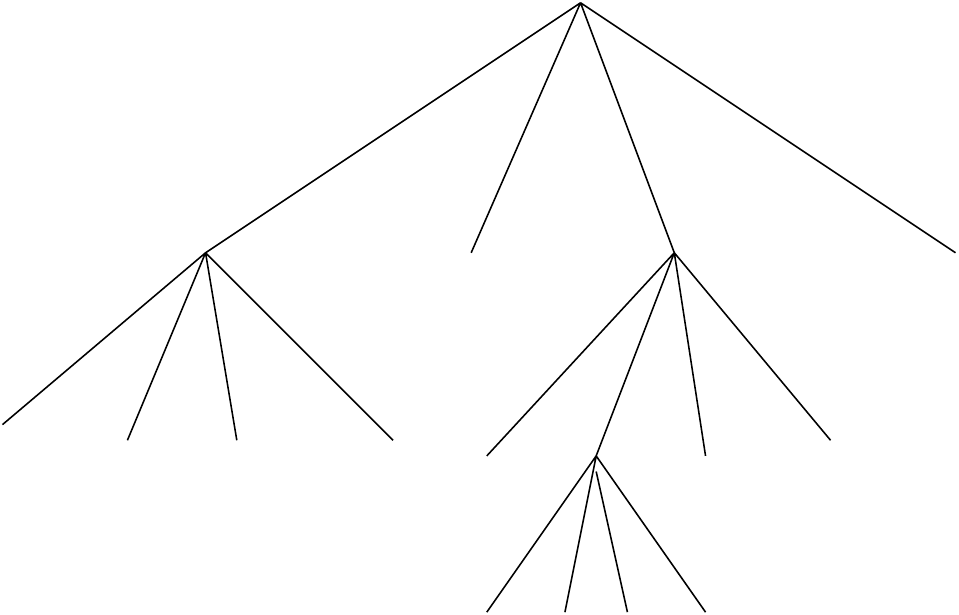}}
\vspace{-.1in}  \caption{The decomposition of a 2-D data
 and its corresponding QT ($m=13$)}\label{quadtree}
\end{figure}
We also denote $Q_m$ as a set of all QTs that have $m$ leaves, and
for given $q\in Q_m$, define
$R_q:\mathcal{T}_N\rightarrow\{1,\cdots,m\}$ as a mapping that maps
each coordinate of the 2-D data to the one of $m$ leaves in $q$
corresponding to the region containing the coordinate. Although the QT decomposition is limited in the sense that it only decomposes the data into quadrants, it still is shown to be effective in many applications since it gives a compact representation of segmentation and is rich enough for capturing local similarity of data.

\subsection{Peano-Hilbert (PH) Curve}\label{subsection: ph curve}

\begin{figure}[h]
\centering \subfigure[PH(4) curve ($16\times16$)]{\label{ph16}
\includegraphics[width=0.2\textwidth]{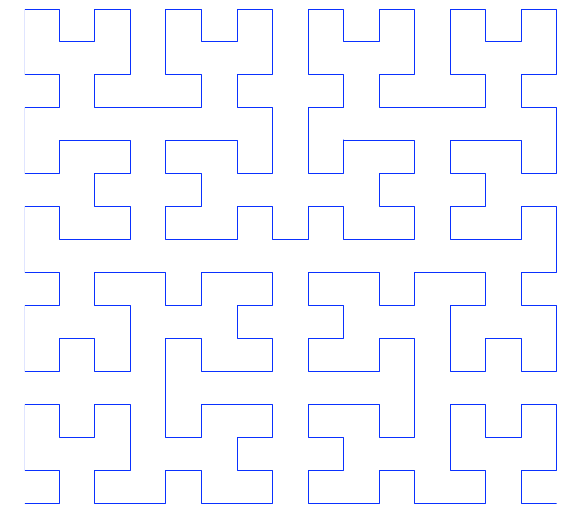}}
\hspace{.2in} \subfigure[PH(8) curve ($256\times256$)]{\label{ph256}
\includegraphics[width=0.2\textwidth]{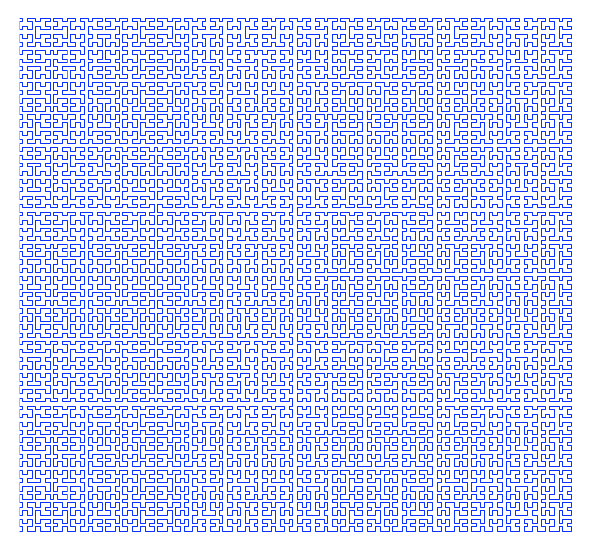}}
\vspace{-.1in}  \caption{Peano-Hilbert curves}\label{ph scan}
\end{figure}

The Peano-Hilbert (PH) curves are well known as space-filling curves. They possess the property that, for any level of quadrants, the curves never leave a quadrant before visiting all the sites within the quadrant. The details of PH curves and scanning orders can be found in \cite[Section II]{LZ86}.  Examples of PH curves for  2-D data with $N=2^4$ and $N=2^8$ are shown in Figure \ref{ph scan}. 

The PH curve naturally defines an ordering of 2-D according to the order in which the PH curve fills the plane.   
Then, for  noisy 2-D data $z^{N\times N}$, we denote $z^n_{\text{PH}}$ as its PH scanned noisy 1-D sequence and $x^n_{\text{PH}}$ as the corresponding clean 1-D sequence. 
%Note that we slightly overloaded the notation that while $z^n$ is a 2-D data but $z^n_{PH}$ is a 1-D data.
In addition, we denote the $i$th index according to the PH scan by $\text{PH}_i$. Note that 
$$
\text{PH}_i\in\mathcal{T}_N, \ \ \ i=1\ldots,n.
$$
 Thus, for example, $z_{\text{PH}_i}$ stands for the 
$i$th component of the $n$-tuple $z_{\text{PH}}^n$, and $\cb_{\text{PH}_i,\text{2D}}\in\Cb_{k,\text{2D}}$ is the 2-D $k$-th order context at that location.

\subsection{Derivation of the scheme and performance guarantees}\label{subsection: main results}

Equipped with the definitions and notation in the previous subsections, we now follow the line of argument paralleling that  in \cite{sdude}:
begin with the case of symbol-by-symbol denoisers  ($k=0$) to build the main idea, then
move on to the general $k$-th order case. For simplicity, we assume
$N=2^{r}$, and $n=N\times N=2^{2r}$.

\subsubsection{Competing with Combinations of Single-Symbol Denoisers  ($k=0$)}\label{subsec: 0th order}

Consider a 2-D $n$ tuple of single-symbol denoisers
$\mathbf{S}=\{s_t:t\in\mathcal{T}_N\}\in\mcS_0^n$. For such
$\mathbf{S}$, we can associate the 2-D $n$-block denoiser
$\hat{\mathbf{X}}^{n,\mathbf{S}}_{\text{2D}}$ as
\begin{equation} \label{eq: *}
\hat{X}_{t,\text{2D}}^{\mathbf{S}}(z^{N\times N})=s_t(z_t)
\end{equation}
  for all
$t\in\mathcal{T}_N$. 
In order to construct the reference class based on QT decomposition described in Section \ref{subsec: motivation}, we define 
a subset
$\mathbf{S}_m(q)\subset\mcS_0^n$ associated with a given QT $q\in Q_m$ as
$$
\mathbf{S}_m(q)\triangleq\{\mathbf{S}\in\mathcal{S}_0^n:s_i=s_j \
\textrm{if} \ R_q(i)=R_q(j),\ \ \text{for all} \ \ i,j\in\mathcal{T}_N\}.
$$
In words, $\mathbf{S}_m(q)$ is a set of 2-D $n$-tuples of single-symbol
denoisers, of which denoising rules are constant within each of the  $m$ distinct regions defined by
a QT $q\in Q_m$. Now, for a fixed $n$ and $m$, define a set
$\mcS_{0,Q_m}^n\subset\mcS_0^n$ as
\begin{align}
\mcS_{0,Q_m}^{n}\triangleq  \bigcup_{q\in Q_m}
\mathbf{S}_m(q) ,\label{ref class 0}
\end{align}
which is a set of all possible configurations of single-symbol denoisers
confined to be constant in regions defined by QTs with $m$ leaves. Following is a simple lemma presenting a lower bound on the size of the set $\mcS_{0,Q_m}^n$ in terms of the number of segmented regions $m$.

\begin{lem}\label{set size}
The set $\mcS_{0,Q_m}^n$ defined in  \eq{ref class 0} satisfies 
$$|\mcS_{0,Q_m}^n|=\Omega\big(3^{\frac{m}{3}}\big). 
$$
%|\mcS|^m\prod_{i=1}^{\frac{m-1}{3}-1}(3i+1)$$
\end{lem}
\emph{Proof:} By the definition of the QT,  we observe that the number of leaves always has the form of $m=3j+1,\  \text{where} \ \ j\in\mathbb{N}\cup\{0\} $. Then, for given $m$, we can see that
\begin{eqnarray}
|\mcS_{0,Q_m}^n|&\geq&\prod_{i=0}^{j-1}(3i+1)\geq 3^{\frac{m-4}{3}},\nonumber%=\Omega\big(e^m\big).\ \ \quad\blacksquare
\end{eqnarray}
which is from the fact that a new segmentation of a quadrant can happen in any of the leaves of the original QT. 
Therefore, $$\lim_{m\rightarrow\infty}\frac{|\mcS_{0,Q_m}^n|}{3^{m/3}}>0,$$ and we have the lemma. $\quad\blacksquare$

Given the reference class of switching single-symbol denoisers (\ref{ref class 0}), we define the performance target for given 2-D data $(x^{N\times N}, z^{N\times N})$ as
\begin{align}
D^{2D}_{0,m}(x^{N\times N}, z^{N\times N})=\min_{\mathbf{S}\in\mcS_{0,Q_m}^n}\frac{1}{n}\sum_{t\in\mathcal{T}_N}\Lambda(x_t,s_t(z_t)),\label{perf target 0}
\end{align}
i.e., the best denoising performance attainable among all combinations of single symbol denoisers in $\mcS_{0,Q_m}^{n}$. 
%Now, as a performance target, we slightly overload the notation of
%the one in \cite{sdude} and define $$
%D^{2D}_{0,m}(x^n,z^n)=\min_{\mathbf{S}\in\mcS_{0,m}^n}\frac{1}{n}\sum_{t\in\mathcal{T}_N}\Lambda(x_t,s_t(z_t)).
%$$
%
%
%
%the minimum normalized loss of 2-D data $(x^n,z^n)$ that can be
%attained by the best combination of single-symbol denoisers among
%$\mcS_{0,m}^n$.
In order to find the combination of single-symbol denoisers that asymptotically achieve (\ref{perf target 0})  based only  on
$Z^{N\times N}$, we may again use the idea of utilizing the estimated loss (\ref{estimated loss}) in
place of the true loss as in \cite{sdude} to find 
\begin{align}
\arg\min_{\mathbf{S}\in\mcS_{0,Q_m}^n}\frac{1}{n}\sum_{t\in\mathcal{T}_N}\ell(Z_t,s_t).\label{min estimated loss 0}
\end{align}
However, the naive brute-force algorithm to find the achiever of (\ref{min estimated loss 0}) requires the exhaustive search over the set $\mcS_{0,Q_m}^n$, which results in \emph{exponential} complexity in $m$ as specified in Lemma \ref{set size}. In contrast to the 1-D case, an efficient algorithm that directly finds the best combination of single-symbol denoisers in $\mcS_{0,Q_m}^n$ does not appear to exist.  To circumvent this issue, the PH scanning comes into play and serves as a key component for devising an efficient algorithm to attain performance essentially at least as good as \eq{perf target 0}. To this end, we define another set of combinations of single-symbol desnoisers
\begin{eqnarray}
\mcS_{0,m}^{\text{PH}(n)}\triangleq\big\{\mathbf{S}\in\mcS_0^n: \sum_{i=1}^n\mathbf{1}_{\{s_{\text{PH}_{i-1}}\neq s_{\text{PH}_i}\}}\leq m\big\}\label{s comb}. 
\end{eqnarray}
 In words, $\mcS_{0,m}^{\text{PH}(n)}$ is a set of combinations of single-symbole denoisers that have at most $m$ switches when the denoisers are ordered according to the PH scanning order. Equation \eq{s comb} is identical to \cite[eq. (20)]{sdude} except that it is for the PH scanned 1-D data of the original 2-D data. We can now define 
\begin{align}
\hat{\mathbf{S}}=\arg\min_{\mathbf{S}\in\mcS_{0,m}^{PH(n)}}\frac{1}{n}\sum_{i=1}^n\ell(Z_{\text{PH}_i},s_{\text{PH}_i}),\label{min estimated loss 1}
\end{align}
which can be found  with \emph{linear} complexity both in $n$ and $m$ by the two-pass dynamic programming algorithm established in \cite[Section V]{sdude}. We denote our 2-D $(0,m)$-S-DUDE as
$\hat{\mathbf{X}}^{n,0,m}_{\textrm{2D univ}}$, and define it to be
$\hat{\mathbf{X}}_{\text{2D}}^{n,\hat{\mathbf{S}}}$ (recalling the notation $\hat{\mathbf{X}}_{\text{2D}}^{n,\mathbf{S}}$ from \eq{eq: *}). Before stating the performance guarantee of our scheme, we have following lemma.

\begin{lem}\label{lem: target lemma}
Define a quantity
$$
D_{0,m}^{\text{PH}}(x^{N\times N},z^{N\times N})\triangleq\min_{\Sb\in\mcS_{0,m}^{\text{PH}(n)}}\frac{1}{n}\sum_{t\in\mathcal{T}_N}\Lambda(x_t,s_t(Z_t)).
$$
Then, we have
$$
D_{0,m}^{\text{PH}}(x^{N\times N},Z^{N\times N})\leq D^{2D}_{0,m}(x^{N\times N},z^{N\times N}).
$$
for every $(x^{N\times N}, z^{N\times N})$.
\end{lem}
\emph{Proof:} From the definitions (\ref{ref class 0}) and (\ref{s comb}), we can see that 
$$
\mcS_{0,Q_m}^n\subseteq\mcS_{0,m}^{PH(n)},
$$
since any 2-D $n$-tuple single-symbol denoisers in $\mcS_{0,Q_m}^n$ would be also in $\mcS_{0,m}^{PH(n)}$ after reordering them in the PH scanning order. This is because PH scan never leaves a quadrant before visiting all the data points in a quadrant, and the shifting positions for sequences in $\mcS_{0,m}^{PH(n)}$ can appear anywhere in the data, whereas the shifting positions in $\mcS_{0,Q_m}^n$ always appear on the boundary of the quadtree decomposed quadrants. Therefore, as the objective functions are identical for $D_{0,m}^{\text{PH}}(x_{\text{PH}}^n,z_{\text{PH}}^n)$ and $D^{\text{2D}}_{0,m}(x^{N\times N},z^{N\times N})$, we get the lemma. $\quad\blacksquare$

The following theorem gives the concentration result on the performance of $\hat{\mathbf{X}}^{n,0,m}_{\textrm{2D univ}}$. 
\begin{theorem}\label{thm: perf guarantee 0}
For $\hat{\mathbf{X}}^{n,0,m}_{\textrm{2D univ}}$ defined in (\ref{min estimated loss 1}), and for all $\epsilon>0$ and $x^{N\times N}\in\mcX^{N\times N}$, we have
\begin{eqnarray}
&&\textrm{\emph{Pr}}\left(L_{\hat{\mathbf{X}}^{n,0,m}_{\textrm{2D univ}}}(x^{N\times N},Z^{N\times N}) - D_{0,m}^{\text{2D}}(x^{N\times N},Z^{N\times N})>\epsilon \right)\nonumber \\
&&\leq 2\exp\Big(-n\Big[\frac{\epsilon^2}{2L_{\max}^2}-2\Big\{h\Big(\frac{m}{n}\Big)+\frac{(m+1)\ln|\mcS|}{n}\Big\}\Big]\Big)\label{prob bound}
\end{eqnarray}
where $h(x) = -x\ln x -(1-x)\ln(1-x)$ for $0\leq x \leq 1$, $L_{\max}=\max_{x\in\mcX, z\in\mcZ}\Lambda(x,z)+\max_{z\in\mcZ, s\in\mcS}\ell(z,s)$, and $|\mcS|=|\hat{\mcX}|^{|\mcZ|}$, the size of the set of single-symbol denoisers. 
\end{theorem}
\emph{Proof:} From the union bound, we have
\begin{eqnarray}
&&\textrm{Pr}\left(L_{\hat{\mathbf{X}}^{n,0,m}_{\textrm{2D univ}}}(x^{N\times N},Z^{N\times N}) - D_{0,m}^{\text{2D}}(x^{N\times N},Z^{N\times N})>\epsilon \right)\nonumber\\
&& = \textrm{Pr}\left(L_{\hat{\mathbf{X}}^{n,0,m}_{\textrm{2D univ}}}(x^{N\times N},Z^{N\times N}) -D_{0,m}^{\text{PH}}(x^{N\times N},Z^{N\times N})+D_{0,m}^{\text{PH}}(x^{N\times N},Z^{N\times N})- D_{0,m}^{\text{2D}}(x^{N\times N},Z^{N\times N})>\epsilon \right)\nonumber\\
&&\leq \textrm{Pr}\left(L_{\hat{\mathbf{X}}^{n,0,m}_{\textrm{2D univ}}}(x^{N\times N},Z^{N\times N}) -D_{0,m}^{\text{PH}}(x^{N\times N},Z^{N\times N})>\epsilon\right),\label{ph prob}
\end{eqnarray} 
since $$\text{Pr}\left(D_{0,m}^{\text{PH}}(x^{N\times N},Z^{N\times N})- D_{0,m}^{\text{2D}}(x^{N\times N},Z^{N\times N})>0\right)=0$$ from Lemma \ref{lem: target lemma}. Therefore, the event in \eq{ph prob} becomes identical to the one in the 1-D problem, and the probability bound (\ref{prob bound}) is obtained by the exactly same analysis given in \cite[Theorem 2]{sdude}. $\quad\blacksquare$

\subsubsection{Competing with Combinations of $k$-th order denoisers}
Establishing the result on the single-symbol denoiser case, we now can move on to the general case of competing with $k$-th order denoisers. For general $k>0$, let $\tilde{k}=\lceil k/4\rceil$ and
$n_k=(N-\tilde{k})\times(N-\tilde{k})$. Define
$$\mathcal{T}_{N_k}\triangleq\{t\in\mathbb{Z}^2:t=(t_1,t_2),
\tilde{k}+1\leq t_1\leq N-\tilde{k}, \tilde{k}+1\leq t_2\leq
N-\tilde{k}\},$$ which is a subset of $\mathcal{T}_N$ with size
$|\mathcal{T}_{N_k}|=n_k$. For given $z^{N\times N}$, we define $n_k$-tuple of
($k$-th order denoiser induced) single-symbol denoisers $$
\mathbf{S}_k(z^{N\times N})\triangleq\{s_{k,t}(\cb_{t,\text{2D}},\cdot)\}_{t\in\mathcal{T}_{N_k}}\in\mcS_0^{n_k},
$$
where $\cb_{t,\text{2D}}$ is the 2-D $k$-th order context defined in Section \ref{subsection: two
dimensional context}.
%
%where, to recall, $s_k(\cb_t,\cdot)$ is the single-symbol denoiser
%induced from $s_{k,t}\in\mcS_k$ and the 2-D context $\cb_t$.
For brevity, we suppress the dependence on $z^{N\times N}$ in
$\mathbf{S}_k(z^{N\times N})$ and denote it as $\mathbf{S}_k$. Similarly as in
the case of $k=0$, for given $\mathbf{S}_k$, we can associate
the 2-D $n$-block denoiser $\hat{\mathbf{X}}_{2\text{D}}^{n,\mathbf{S}_k}$ as
$$\hat{\mathbf{X}}_{t,\text{2D}}^{\mathbf{S}_k}(z^{N\times N})=s_{k,t}(\cb_{t,\text{2D}},z_t) \ \ \ \ t\in\mathcal{T}_{N_k}. $$  
%We denote the normalized cumulative loss as
%$$
%L_{\hat{\mathbf{X}}^{n,\mathbf{S}_k}}(x^n,Z^n)=\frac{1}{n_k}\sum_{t\in\mathcal{T}_{N_k}}\Lambda(x_t,s_{k,t}(\cb_t,z_t))
%$$
%and the estimated normalized cumulative loss as
%$$\tilde{L}_{\mathbf{S}_k}(z^n)=\frac{1}{n_k}\sum_{t\in\mathcal{T}_{N_k}}\ell(z_t,s_{k,t}(\cb_t,\cdot)).$$
%Then, we have following lemma that generalizes Lemma \ref{lem:
%concentration}, and parallels \cite[Lemma 2]{sdude}.
%
%\begin{lem} Fix $\epsilon>0$. For any fixed sequence
%$\{s_{k,t}\}_{t\in\mathcal{T}_{N_k}}$ and all $x^n\in\mcX^n$,
%\begin{align}
%\emph{Pr}\Big(L_{\hat{\mathbf{X}}^{n,\mathbf{S}_k}}(x^n,Z^n)-\tilde{L}_{\mathbf{S}_k}(Z^n)>\epsilon\Big)\leq
%(\tilde{k}+1)^2\exp\Big(-\frac{2n_k\epsilon^2}{(\tilde{k}+1)^2L_{\max}^2}\Big)
%\end{align}
%\end{lem}
%\emph{Proof :} The proof of this lemma essentially is similar to the
%one in \cite[Lemma 2]{sdude}. We can first define a set of sub-data
%indices
%$$
%\mathcal{I}_{d_1,d_2}\triangleq\{t:t\in\mathcal{T}_k, \ \ t_1\equiv
%d_1\mod (\tilde{k}+1),\ \ t_2\equiv d_2\mod(\tilde{k}+1)\}
%$$
%for $0\leq d_1\leq\tilde{k}$, $0\leq d_2\leq\tilde{k}$. Then,
%conditioned on all the noisy data whose indices are not in
%$\mathcal{I}_{d_1,d_2}$, we observe that
%$\{Z_t:t\in\mathcal{I}_{d_1,d_2}\}$ are conditionally independent.
%By applying the concentration inequality of Lemma \ref{lem:
%concentration} on each subsequence, and optimizing constants as in
%\cite[Lemma 2]{sdude}, we can obtain the lemma. $\quad\blacksquare$
As in the 1-D case \cite{sdude}, for each $\cb\in\Cb_{k,2\text{D}}$, we 
define $\mathcal{T}(\cb)=\{\tau:\cb_{\tau,\text{2D}}=\cb,
\tau\in\mathcal{T}_{N_k}\}$ as the set of indices of $z^{N\times N}$ where the
2-D context equals $\cb$. Then, for fixed $m$, each quadtree $q\in Q_m$, and given $z^{N\times N}$,
define a subset $\mathbf{S}_{k,m}(q)\subset\mcS_0^{n_k}$ as
\begin{eqnarray}
\mathbf{S}_{k,m}(q)\triangleq \bigcup_{\cb\in\Cb_{k,\text{2D}}}\big\{\{s_{k,t}(\cb,\cdot)\}_{t\in\mathcal{T}(\cb)}:
s_{k,i}(\cb,\cdot)=s_{k,j}(\cb,\cdot)\ \textrm{if} \
R_q(i)=R_q(j)\big\} \nonumber,
%
%
%
%s_{k,i}(\cb,\cdot)=s_{k,j}(\cb,\cdot)\nonumber\\
%\ \ \ \ \ \ \ \ \ \ \ \ \ \ \ \ \ \ \textrm{if} \ R_q(i)=R_q(j),\
%\forall\ i,j\in\mathcal{T}(\cb)\}.
\end{eqnarray}
the set of $n_k$-tuples of ($k$-th order denoiser induced)
single-symbol denoisers that, within the sub-data points
$\{t:t\in\mathcal{T}(\cb)\}$ for each $\cb\in\Cb_{k,\text{2D}}$, are identical
on the regions decomposed by $q\in Q_m$. Then, define
$$
\mcS_{k,Q_m}^n(z^{N\times N})= \bigcup_{q\in
Q_m}\mathbf{S}_{k,m}(q) 
$$
as all  combinations of ($k$-th order denoiser induced)
single-symbol denoisers that, within each subsequence associated
with each $\cb\in\Cb_{k,\text{2D}}$, is confined to remain constant within each of  the
regions determined by QTs with $m$ leaves. Again, for brevity, the
dependence on $z^{N\times N}$ in $\mcS_{k,Q_m}^n(z^{N\times N})$ is suppressed, and we
write  $\mcS_{k,Q_m}^n$.
%This is also an overloaded notation of
%that in \cite{sdude}.
Note that the above notation simply generalizes that of Section \ref{subsec: 0th order} by parallelizing over each 2-D context. Finally, in analogy to (\ref{perf target 0}), for given $(x^{N\times N}, z^{N\times N})$, we define the $k$-th order performance target as 
\begin{eqnarray}
D_{k,m}^{\text{2D}}(x^{N\times N}, z^{N\times N})=\min_{\mathbf{S}\in\mcS_{k,Q_m}^n}\frac{1}{n_k}\sum_{t\in\mathcal{T}_{N_k}}\Lambda(x_t,s_{k,t}(\cb_{t,\text{2D}}, z_t)).\label{perf target k}
\end{eqnarray}
%the minimum normalized cumulative loss of $(x^n,z^n)$ that can be
%achieved by the best combination in $\mcS_{k,m}^n$.
Here too, using the estimated loss to directly find the combination of $k$-th order 2-D sliding window denoisers in $\mcS_{k,Q_m}^n$ that achieves (\ref{perf target k}) may require prohibitive complexity in $m$. Therefore, as in Section \ref{subsec: 0th order}, we  consider the 2-D data in the order of PH scanning, this time independently for each subsequence defined by each 2-D context $\cb\in\Cb_{k,\text{2D}}$. To that end, we define $\mcS_{k,m}^{\text{PH}(n)}$ that parallels  (\ref{s comb}) and \cite[(28)]{sdude} as
\begin{eqnarray}
\mcS_{k,m}^{\text{PH}(n)} = \{\mathbf{S}_k: \{s_{k,\tau}(\cb,\cdot)\}_{\tau\in\mathcal{T}(\cb)}\in\mcS_{0,m(\cb)}^{\text{PH}(n(\cb))}\ \ \text{for all} \ \ \cb\in\Cb_{k,\text{2D}}\},\label{ph scan k}
\end{eqnarray}
where $n(\cb)= |\mathcal{T}(\cb)|$ and $m(\cb)=\min\{m,n(\cb)\}$. Although it may look complex, (\ref{ph scan k}) is simply a set of combination of $k$-th order sliding window denoisers that shift at most $m$ times along the PH-scanned subsequence for each 2-D context $\cb\in\Cb_{k,\text{2D}}$. With this notation and definitions, we  define
\begin{eqnarray}
\hat{\mathbf{S}}_{k,m} = \arg\min_{\mathbf{S}_k\in\mcS_{k,m}^{\text{PH}(n)}}\frac{1}{n_k}\sum_{\{i:\text{PH}_i\in\mathcal{T}_{N_k}\}}\ell(Z_{\text{PH}_i}, s_{k,\text{PH}_i}(\cb_{\text{PH}_i,\text{2D}},\cdot))\label{min estimated loss k}
\end{eqnarray}
and the 2-D $(k,m)$-S-DUDE, $\hat{\mathbf{X}}^{n,k,m}_{\textrm{2-D
univ}}$, as $\hat{\mathbf{X}}_{\text{2D}}^{n,\hat{\mathbf{S}}_{k,m}}$. By applying the dynamic programming algorithm of the 1-D case \cite{sdude} for each PH-scanned subsequence defined by each context $\cb\in\Cb_{k,\text{2D}}$, we can find (\ref{min estimated loss k}) with complexity linear in both $n$ and $m$. A subtle point to emphasize here is that although we apply the 1-D scheme on the PH scanned 1-D data, the subsequences  that we apply our algorithm on are defined by the 2-D $k$-th order contexts. In this way, our scheme still competes with all combinations of 2-D $k$-th order sliding window denoisers with high probability, as is established in the following result.% gives the performance guarantee of our scheme. 

%

%we once more consider the PH-scanned sequence of the 2-D data and define $\mcS_{k,m}^{\text{PH}(n)}$, the combination of $k$-order sliding window denoisers that shifts at most $m$ times along the PH-scanned sequence for each 2-D context $\mathbf{C}_{k,\text{2D}}$, that parallels (\ref{s comb}) and \cite[(28)]{sdude}. 

%

%

%To attain this performance target only based on $Z^{N\times N}$, we again use
%the estimated loss and define
%\begin{eqnarray}
%\hat{\mathbf{S}}_{k,m}=\arg\min_{\mathbf{S}\in\mcS_{k,m}^n}\frac{1}{n_k}\sum_{t\in\mathcal{T}_{N_k}}\ell(z_t,s_{k,t}(\cb_t,\cdot)),\label{k
%sdude def}
%\end{eqnarray}
%and the 2-D $(k,m)$-S-DUDE, $\hat{\mathbf{X}}^{n,k,m}_{\textrm{2-D
%univ}}$, as $\hat{\mathbf{X}}^{n,\hat{\mathbf{S}}_{k,m}}$. Then, we
%can prove the following theorem, asserting that the loss of
%$\hat{\mathbf{X}}^{n,k,m}_{\textrm{2-D univ}}$ will be close to
%$D_{k,m}(x^n,Z^n)$ with high probability.
%%
%%
%%building on the similar lemma as Lemma \ref{lem: concentration}, we
%%have following theorem generalizing Theorem \ref{thm: single-symbol
%%case}, whose proof will be found in \cite{sdude2D}.

%%\begin{lem}
%%For all 2-D data $(x^n,z^n)$,
%%$$
%%D_{k,m}(x_{PH}^n,z_{PH}^n)\leq D_{k,m}^{2D}(x^{N\times N},z^{N\times N}).
%%$$
%%\end{lem}

%
%$$
%D_{k,m}^{2D}(x^{N\times N},z^{N\times N})=\min_{\mathbf{S}\in\mcS_{k,m}^n}\frac{1}{n_k}\sum_{t\in\mathcal{T}_{N_k}}\Lambda(x_t,s_{k,t}(\cb_t,z_t)).
%$$

%$$
%D_{k,m}^{PH}
%$$

%

\begin{theorem}\label{thm: perf guarantee k}
For all $\epsilon>0$ and $x^{N\times N}\in\mcX^{N\times N}$,
\begin{eqnarray}
& &\textrm{\emph{Pr}}\Big(L_{\hat{\mathbf{X}}^{n,k,m}_{\textrm{2-D
univ}}}(x^{N\times N},Z^{N\times N})-D_{k,m}^{\text{2D}}(x^{N\times N},Z^{N\times N})>\epsilon\Big)\nonumber\\
&\leq&2(\tilde{k}+1)^2\exp\Big(-n_k\cdot\Big[\frac{(\epsilon/L_{\max})^2}{2(\tilde{k}+1)^2}-2|\mcZ|^{2k}\cdot
\Big\{h\Big(\frac{m}{n_k}\Big)+\frac{(m+1)\ln |\mcS|}{n_k}\Big\}\Big]\Big).\label{eq: thm4}
\end{eqnarray}
%where $|\mcS|=|\mcZ|^{|\hat{\mcX}|}$ and
%$L_{\max}=\Lambda_{\max}+\ell_{\max}$.
\end{theorem}
\emph{Proof :} Once we define 
$$
D_{k,m}^{\text{PH}}(x^{N\times N},z^{N\times N})=\min_{\mathbf{S}_k\in\mcS_{k,m}^{\text{PH}(n)}}\frac{1}{n_k}\sum_{\{i:\text{PH}_i\in\mathcal{T}_{N_k}\}}\Lambda(Z_{\text{PH}_i}, s_{k,\text{PH}_i}(\cb_{\text{PH}_i,\text{2D}},\cdot)),
$$
we can again easily see that 
$$
D_{k,m}^{\text{PH}}(x^{N\times N},z^{N\times N})\leq D_{k,m}^{\text{2D}}(x^{N\times N},z^{N\times N})
$$
for all $(x^{N\times N},z^{N\times N})$ since $\mcS_{k,m}^{\text{PH}(n)}$ is a larger set than $\mcS_{k,Q_m}^n$. Hence, proving the theorem becomes showing
\begin{eqnarray}
& &\textrm{Pr}\Big(L_{\hat{\mathbf{X}}^{n,k,m}_{\textrm{2-D
univ}}}(x^{N\times N},Z^{N\times N})-D_{k,m}(x^{n}_{\text{PH}},Z^{n}_{\text{PH}})>\epsilon\Big)\leq  (\ref{eq: thm4}),\nonumber
\end{eqnarray}
which can be derived by the identical argument as in \cite[Theorem 3]{sdude}.$\quad\blacksquare$

The following result, which is a direct consequence of the above
theorem, can be considered the analogue of Theorem \ref{sdude main}(a)
to 2-D data. It shows that our algorithm is still universal, i.e., regardless of the underlying data, our algorithm asymptotically attains the optimum performance in the reference class.

%
% Note that when $k$ and $m$ grows sufficiently slowly compared
%to $n$, e.g., $k=c_1\log n$ with $c_1<\frac{1}{2\log |\mcZ|}$ and
%$m=c_2\log n$, then, (\ref{eq: thm4}) will be decaying almost
%exponentially, thus summable in $n$. This growth rate of $k$ and $m$
%will ensure the following theorem, which parallels the results of
%\cite{Dude}\cite{sdude}.
\begin{theorem}\label{thm: main theorem}
Suppose $k=k_n$ and $m=m_n$ are such that (\ref{eq: thm4}) is
summable in $n$, e.g., $k=c_1\log n$ with $c_1<\frac{1}{2\log
|\mcZ|}$ and $m=n^{\alpha}$ with $\alpha<1$.  Then, for all
$\mathbf{x}\in\mcX^{\infty \times \infty}$, the sequence of denoisers
$\{\hat{\mathbf{X}}^{n,k,m}_{\textrm{2D univ}}\}$ satisfies
$$
\lim_{N \rightarrow\infty}\big[L_{\hat{\mathbf{X}}^{n,k,m}_{\textrm{2D
univ}}}(x^{N\times N},Z^{N\times N})-D_{k,m}^{\text{2D}}(x^{N\times N},Z^{N\times N})\big]=0\ \textrm{a.s.}
$$
\end{theorem}
\vspace{.02in} \emph{Proof :} The proof combines the summability condition, the Borel-Cantelli lemma, the bound \eq{eq: thm4}, and simple use of the union bound, similarly as in the proof of Theorem \ref{sdude main} in \cite{sdude}. $\quad\blacksquare$\\

\subsection{Algorithm and complexity}\label{subsection: algorithm}
We have shown that by applying the 1-D S-DUDE algorithm in \cite[Section V-A]{sdude} separately on the PH scanned subsequences for each 2-D context $\cb\in\Cb_k$,  the resulting scheme can attain the performance of the best combination of the $k$-th order denoisers that shifts across the $m$ separate quadtree decomposed regions. The pseudo-algorithm for our 2-D S-DUDE is given below:
\begin{algorithm}[H]
\caption{The two-dimensional (2-D) $(k,m)$-Shifting DUDE}
\label{alg1}
\begin{algorithmic}
\REQUIRE $LM_t\in\mathbb{R}^{(m+1)\times |\mcS|}$, $IM_t\in\mathbb{R}^{|\mcS|}$ for $t\in\mathcal{T}_k$, $T\in\mathbb{R}^{|\Cb_k|}, r\in\mathbb{R}^{|\Cb_k|}, q\in\mathbb{R}^{|\Cb_k|}, L\in\mathbb{R}$ as in \cite[Section V-A]{sdude}
%Same memory requirement as in \cite[Section V-A]{sdude}
%$LM_t(i,j)\in\mathbb{R}^{I\times J}$, $IM_t(i)\in\mathbb{R}^I$, \quad$k+1\leq t \leq
%n-k, 1\leq i\leq I, 1\leq j \leq J$,\quad\\
%\ \ \ \ \ \ \ \ \ \ \ $T\in\mathbb{R}^{D}$,$r\in\mathbb{R}^{D}, q\in\mathbb{R}^D$,
%$L\in\mathbb{R}$ 
\ENSURE
$\hat{\Sb}_{k,m}=\{\hat{s}_{k,t}(\cb_{t,\text{2D}},\cdot)\}_{t\in\mathcal{T}_k}$ in \eq{min estimated loss k} and the denoised output $\{\hat{x}_t\}_{t\in\mathcal{T}_k}$
\FOR{increasing order of PH scanned index $\{i:\text{PH}_i\in\mathcal{T}_k\}$}
\STATE identify the 2-D context $\cb_{\text{PH}_i,\text{2D}}$
\STATE run the forward recursion of 1D S-DUDE on PH scanned points $\{\text{PH}_j:\text{PH}_j\in\mathcal{T}(\cb_{\text{PH}_i,\text{2D}}), \ \ j \leq i\}$
\ENDFOR
\FOR{decreasing order of PH scanned index $\{i:\text{PH}_i\in\mathcal{T}_k\}$}
\STATE identify the 2-D context $\cb_{\text{PH}_i,\text{2D}}$
\STATE run the backward recursion of 1D S-DUDE on PH scanned points $\{\text{PH}_j:\text{PH}_j\in\mathcal{T}(\cb_{\text{PH}_i,\text{2D}}), \ \ j \geq i\}$
\STATE identify the best single-symbol denoiser $\hat{s}_t(\cb_{\text{PH},\text{2D}}, \cdot)$ for location $\text{PH}_i$
\STATE obtain the denoised symbol $\hat{x}_{\text{PH}_i} = \hat{s}_t(\cb_{\text{PH},\text{2D}}, z_{\text{PH}_i})$
\ENDFOR
\end{algorithmic}
\end{algorithm}
Note that the PH scanning of the data and running the 1-D S-DUDE on those scanned points can be done simultaneously, not separately. Therefore, the time and memory complexities of our algorithm are exactly the same as those of 1-D S-DUDE : $O(nm )$. Hence, competing with the QT decomposed, shifting 2-D sliding window denoisers for 2-D data is no harder than competing with shifting sliding window denoisers  for 1-D data. 

\subsection{Remarks}
Before presenting the experimental results of our scheme, we have a few remarks regarding our algorithm and analyses in above subsections. 
\begin{enumerate}
\item The performance guarantee results on expected (rather than actual) loss,  and on a stochastic setting where the noise-free image, rather than an individual data array, is a random field, can be derived similarly as in the settings of \cite{Dude} and \cite{sdude}, once equipped with the above semi-stochastic setting result in Theorem \ref{thm: main theorem}. We omit the exercise for conciseness and to refrain from repetition. 

\item It may also be natural to conceive of a denoising algorithm that heuristically finds a QT decomposition by greedily merging child nodes as in \cite{quadtreeFeder}, using the estimated loss. This scheme is practical, and may be competitive with the best shifting sliding window denoisers based on QT decomposition, but is difficult to analyze and obtain rigorous performance guarantees. On the other hand, as the results above guarantee, our scheme, achieves the best possible performance among all scheme in the same reference class, and is practically implementable. 

\item The two components of our scheme, namely, QT decomposition and PH scanning, have been developed independently in previous literature, but in the denoising setting, we see that the marriage of the two is natural since they play complementary roles; the former efficiently segments data points that have similar characteristics and the latter unfolds the 2-D data into 1-D data while preserving the local similarity attained by the former. 

\item Although our algorithm and analysis pertained exclusively to 2-D data case, it is not hard to extend our scheme to the multi-dimensional data case beyond 2-D case, since analogously defining QT decomposition and PH scanning for multi-dimensional data is straightforward. 

\end{enumerate}

\section{Experimental Results}\label{sec: experiments}
As shown above, our 2-D S-DUDE enjoys considerable performance guarantees and efficient implementation, however, it is not clear how effective it might be in practice to compete with the reference class of QT decomposition-based shifting sliding window denoisers. Hence, we show the performance of our scheme on three different sample images and compare with baseline schemes to highlight when use of our algorithm could be advantageous. 
%

%In the previous section, we have described our algorithm and performance guarantees. We now show the effectiveness of 
%combining the PH scan with quadtree decomposition for denoising 2-D data. 

\subsection{Synthetic test image}\label{subsec: synthetic}
\begin{figure}[h]
\centering \subfigure[Clean image]{\label{eysb clean}
\includegraphics[width=0.3\textwidth]{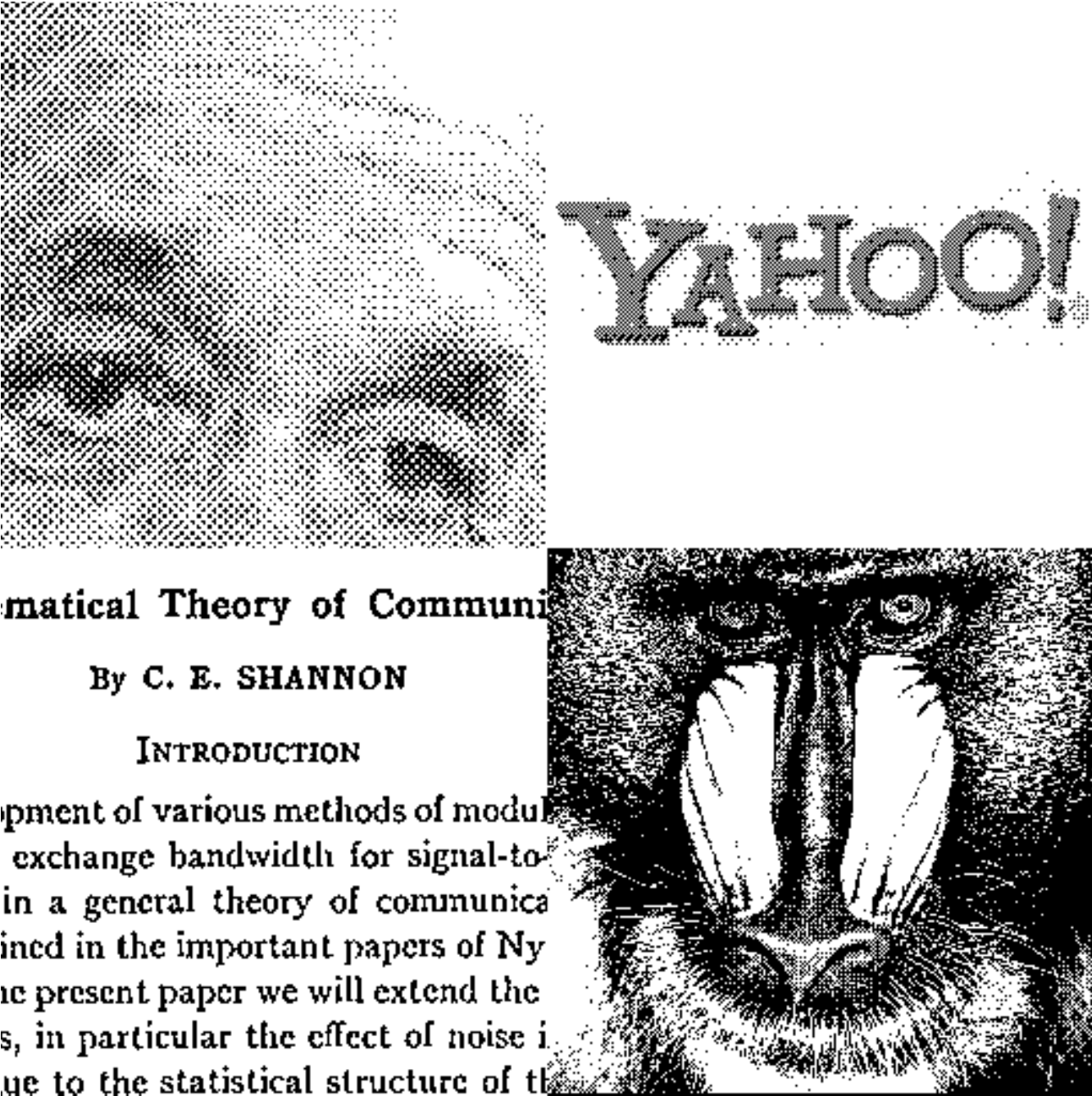}}
\hspace{.2in} \subfigure[Noisy image]{\label{eysb noisy}
\includegraphics[width=0.3\textwidth]{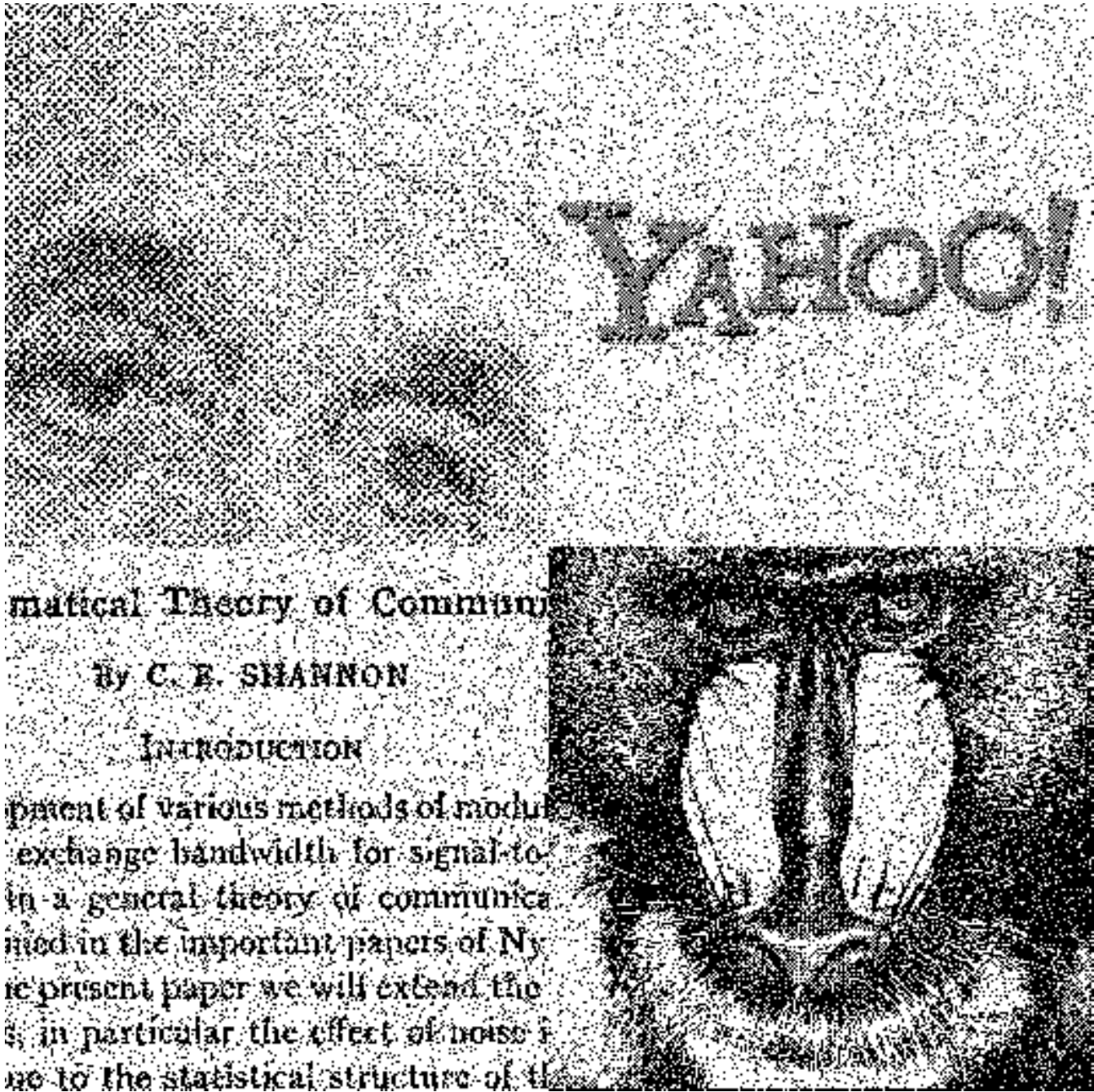}}
\hspace{.2in} \subfigure[Raster scan+1D S-DUDE ($m=4$) ]{\label{eysb 1d sdude}
\includegraphics[width=0.3\textwidth]{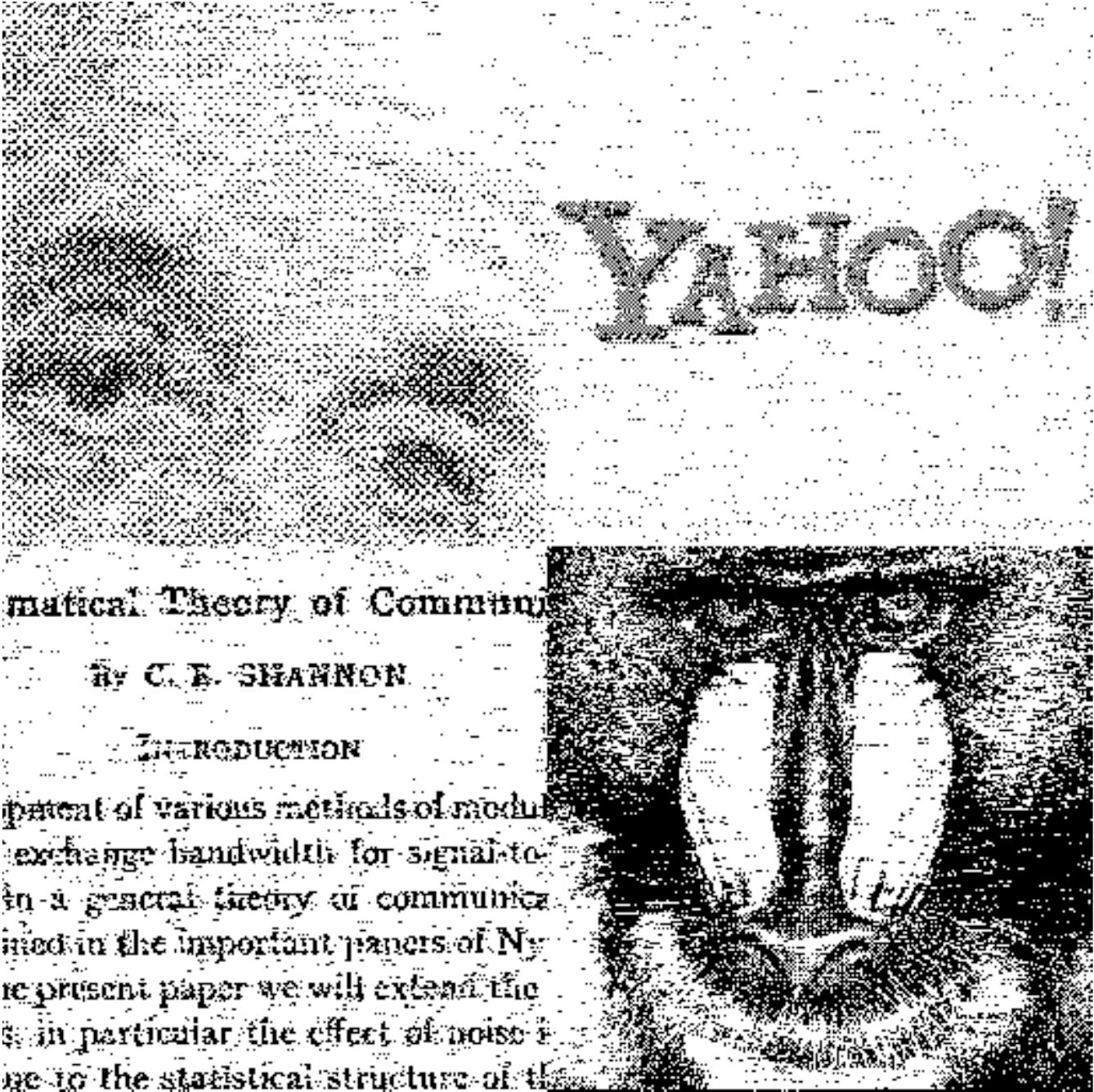}}
\hspace{.2in} \subfigure[2D S-DUDE ($m=4$)]{\label{eysb 2d sdude}
\includegraphics[width=0.3\textwidth]{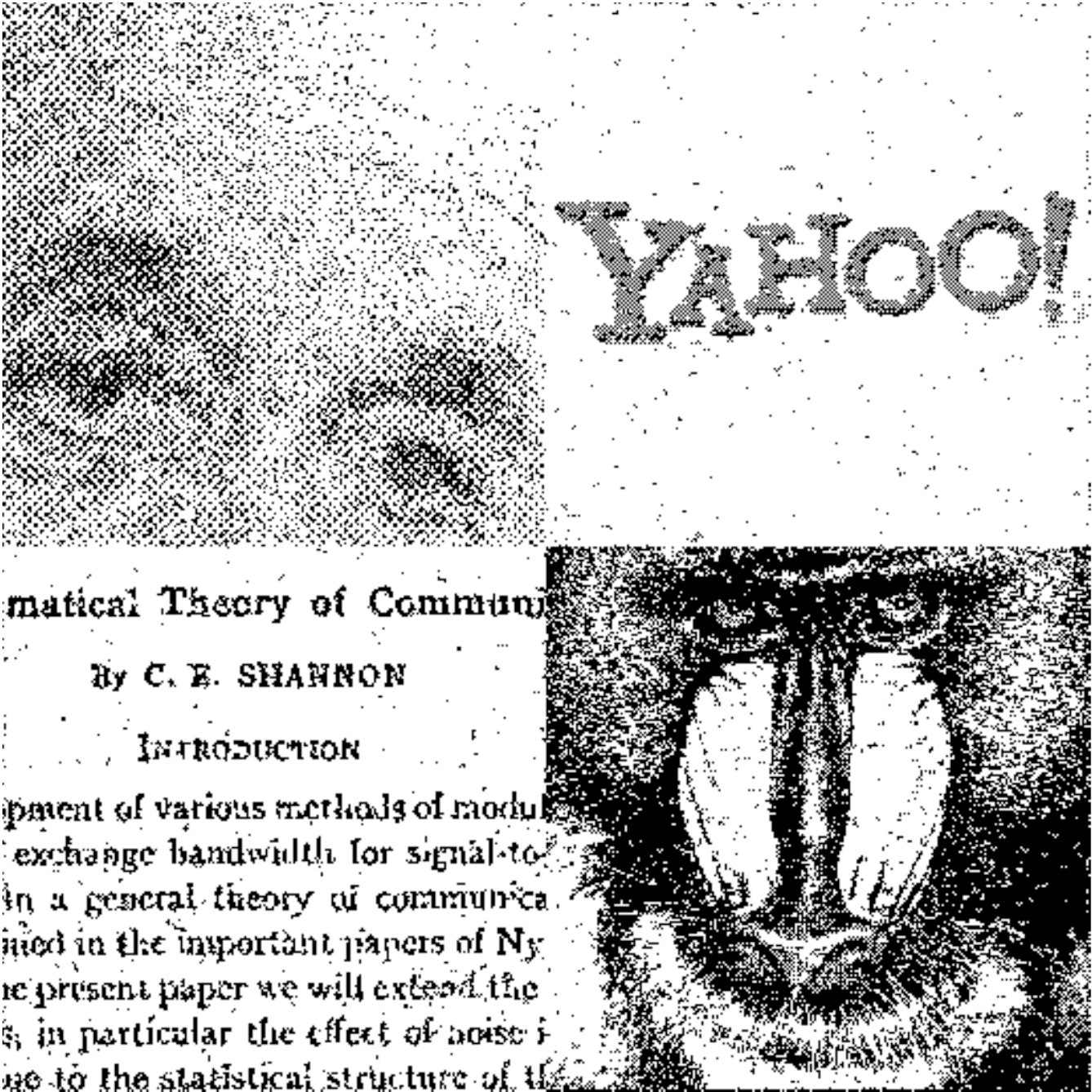}}
\vspace{-.1in}  \caption{Synthetic test image and denoising results}\label{eysb}
\end{figure}

First, we show the denoising results for a synthetic image that showcase the benefit of our 2-D S-DUDE scheme. Figure \ref{eysb clean} shows the clean image that was constructed by pasting four binary sub-images that have different characteristics, and Figure \ref{eysb noisy} is the noisy counterpart corrupted by a binary symmetric channel (BSC) with crossover probability $\delta=0.1$. The total image size is $512\times512$ and each sub-image has size $256\times256$.
We compare our scheme with three different baselines; 
\begin{enumerate}
\item[(i)] \textbf{2-D DUDE} : This simply generalizes the 1-D DUDE in \cite{Dude} to the 2-D data case with using the 2-D contexts defined in Section \ref{subsection: two dimensional context}. Note that this scheme is already superior  to many  state-of-the-art image denoising schemes, as reported in \cite[Section VIII-C]{Dude} and \cite{2d_dude}. It has a single parameter $k$, the context size for the 2-D context. 
\item[(ii)] \textbf{1-D S-DUDE after raster scanning the data}: This scheme first does the simple horizontal raster scan of the image, then applies the 1-D S-DUDE developed in \cite{sdude} on the resulting 1-D sequence. It is the scheme used in \cite[Section V-A]{sdude} for image denoising experiments. The scheme has two parameters - $k$ for the context size for the 1-D context and $m$ for the number of shifts. 
\item[(iii)] \textbf{1-D DUDE after raster scanning the data}:  This scheme is a baseline, which coincides with the scheme in (ii) when the number of shifts $m$ is set to 0. In \cite[Section V-A]{sdude}, (ii) was shown to be superior to this scheme for images with characteristics that are abruptly changing. 
\end{enumerate}
One may think that the only difference between our 2-D S-DUDE and scheme (ii) is that we use the PH scan in place of the raster scan, but there is also a subtle difference that we consider the subsequence points with respect to the 2-D contexts whereas (ii) simply considers the 1-D contexts from the raster-scanned 1-D sequence. 

\begin{figure}[h]
\centering 
\includegraphics[width=0.5\textwidth]{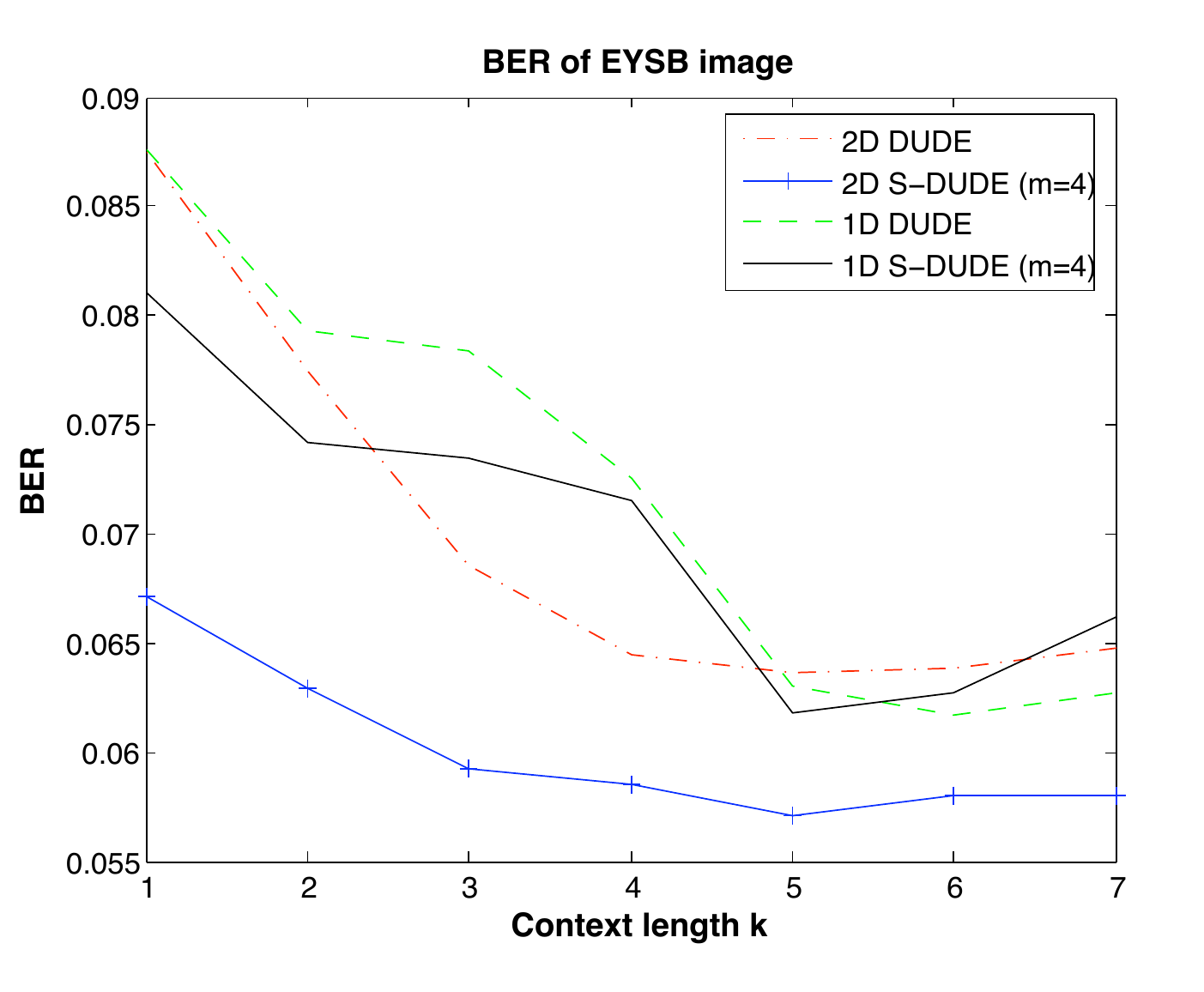}
\caption{Bit error rate comparison for the synthetic image}\label{ber eysb plot}
\end{figure}

Figure \ref{ber eysb plot} shows the bit error rate (BER) results for the four schemes. For (ii) and 2-D S-DUDE, we only show the results for the best $m$ value, which happened to be $m=4$ for both schemes. First, we can see that the difference between (ii) and (iii) is small. This is well expected since we can easily notice that the characteristics of the raster-scanned 1-D sequence of Figure \ref{eysb clean} vary linearly with  respect to the sequence length $n$, which violates the necessary condition on the number of shifts $m$ for (ii) to work, i.e., $m$ should be sublinear in $n$ as specified in \cite[Theorem 5]{sdude}. Second, interestingly, there is also no big difference between (i), which uses the 2-D context,  and (ii) and (iii), which use the 1-D context from the raster-scanned sequence. Probably this shows that for our image, the 1-D contexts from the raster-scanned sequence are enough for capturing the locality of images, but we do not know whether this would be a general phenomenon. 
Finally, we can observe that  our 2-D S-DUDE with $m=4$ clearly dominates all three baseline schemes. Note that, by construction, it would be optimal to first decompose the image in Figure \ref{eysb noisy} into four separate quadrants (quadtree with $4$ leaves) and apply four independent denoisers in each region.  We see that by considering the noisy pixels in the order of PH scanning, our scheme, which knows nothing about the underlying clean image, successfully learns the decomposition of the image and further reduces the BER for denoising compared to other baseline schemes. 

The resulting denoised images for scheme (ii) and our 2-D S-DUDE are shown in Figure \ref{eysb 1d sdude} and Figure \ref{eysb 2d sdude}, respectively.  In line with the BER plot in Figure \ref{ber eysb plot}, we visually see that 2-D S-DUDE is superior to the scheme (ii) not only in terms of the number of errors, but also in terms of detecting the boundaries of images and preserving the textures. Particularly, the texts in Figure \ref{eysb 2d sdude} are more readable and the boundary between the Einstein and  Yahoo! images is more clearly captured in Figure \ref{eysb 2d sdude}.

\subsection{Scanned magazine image}\label{subsec: magazine}
Although the result on the synthetic image is encouraging, one may suspect the image was constructed in favor of 2-D S-DUDE, since it was divided into  different sub-images corresponding to QT quadrants.  We thus test our algorithm on a real image. 
\begin{figure}[h]
\centering \subfigure[Clean image]{\label{magazine clean}
\includegraphics[width=0.3\textwidth]{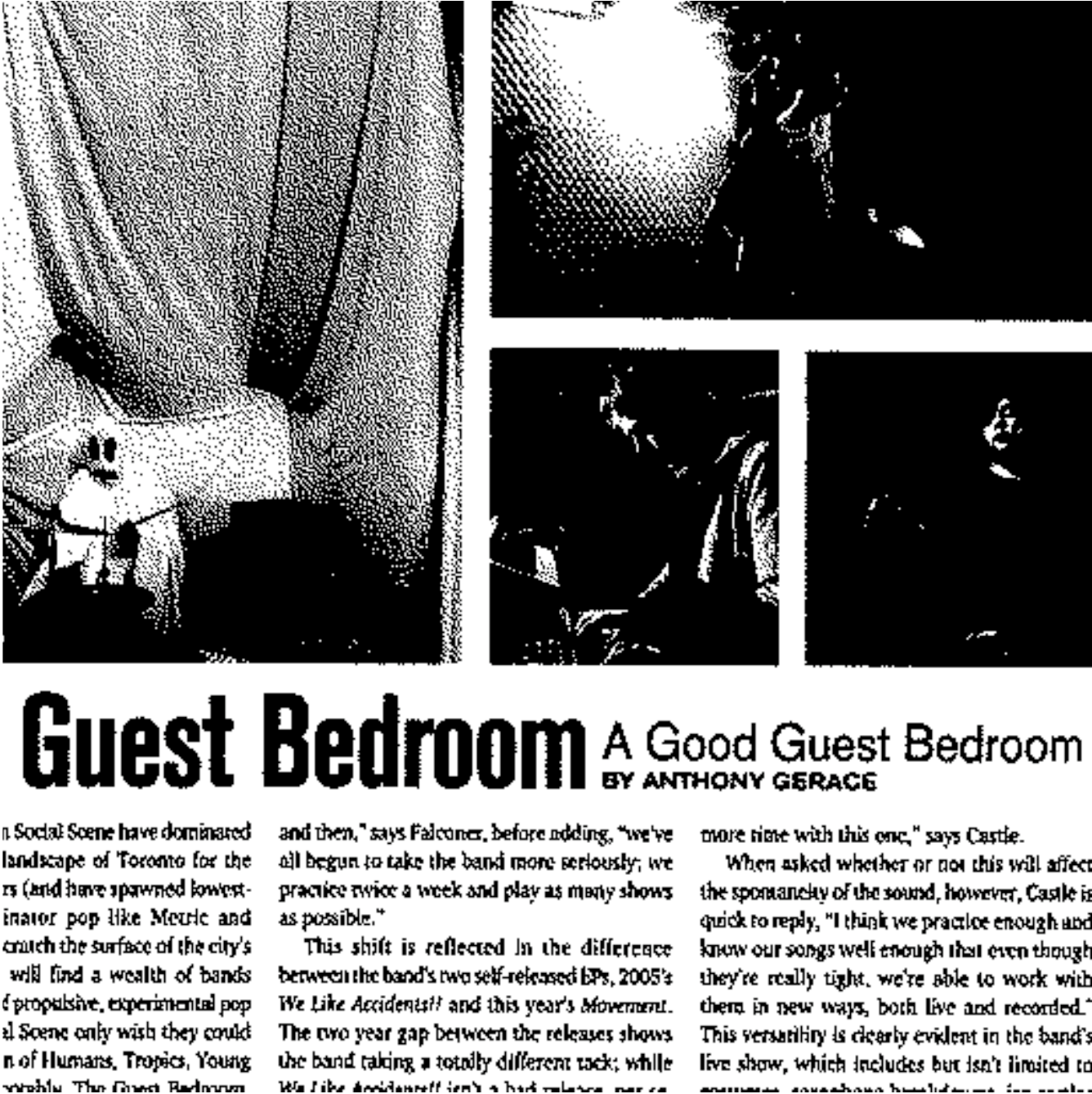}}
\hspace{.2in} \subfigure[Noisy image]{\label{magazine noisy}
\includegraphics[width=0.3\textwidth]{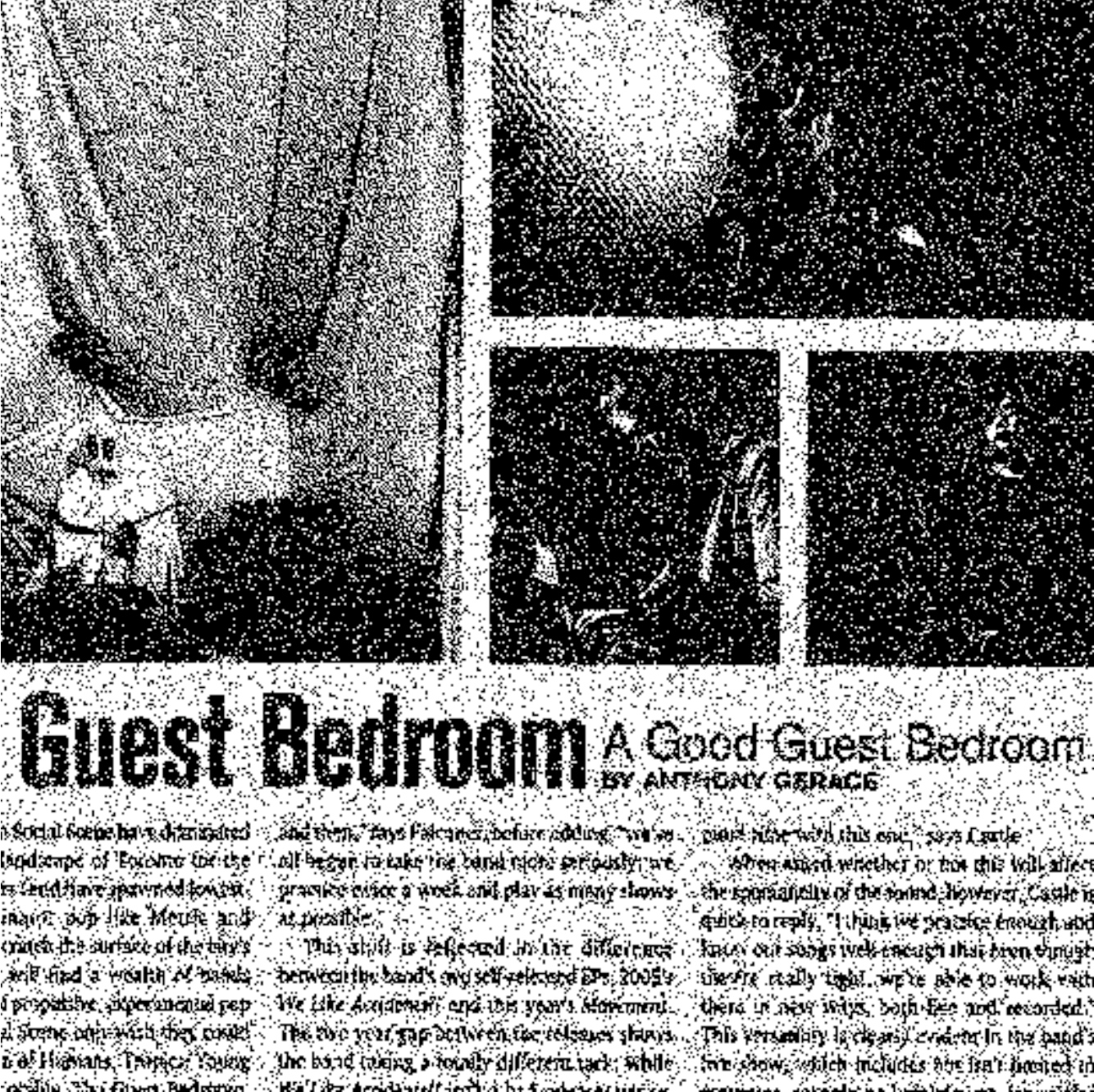}}
\hspace{.2in}  \subfigure[2D DUDE ($k=8$)]{\label{magazine 2d dude}
\includegraphics[width=0.3\textwidth]{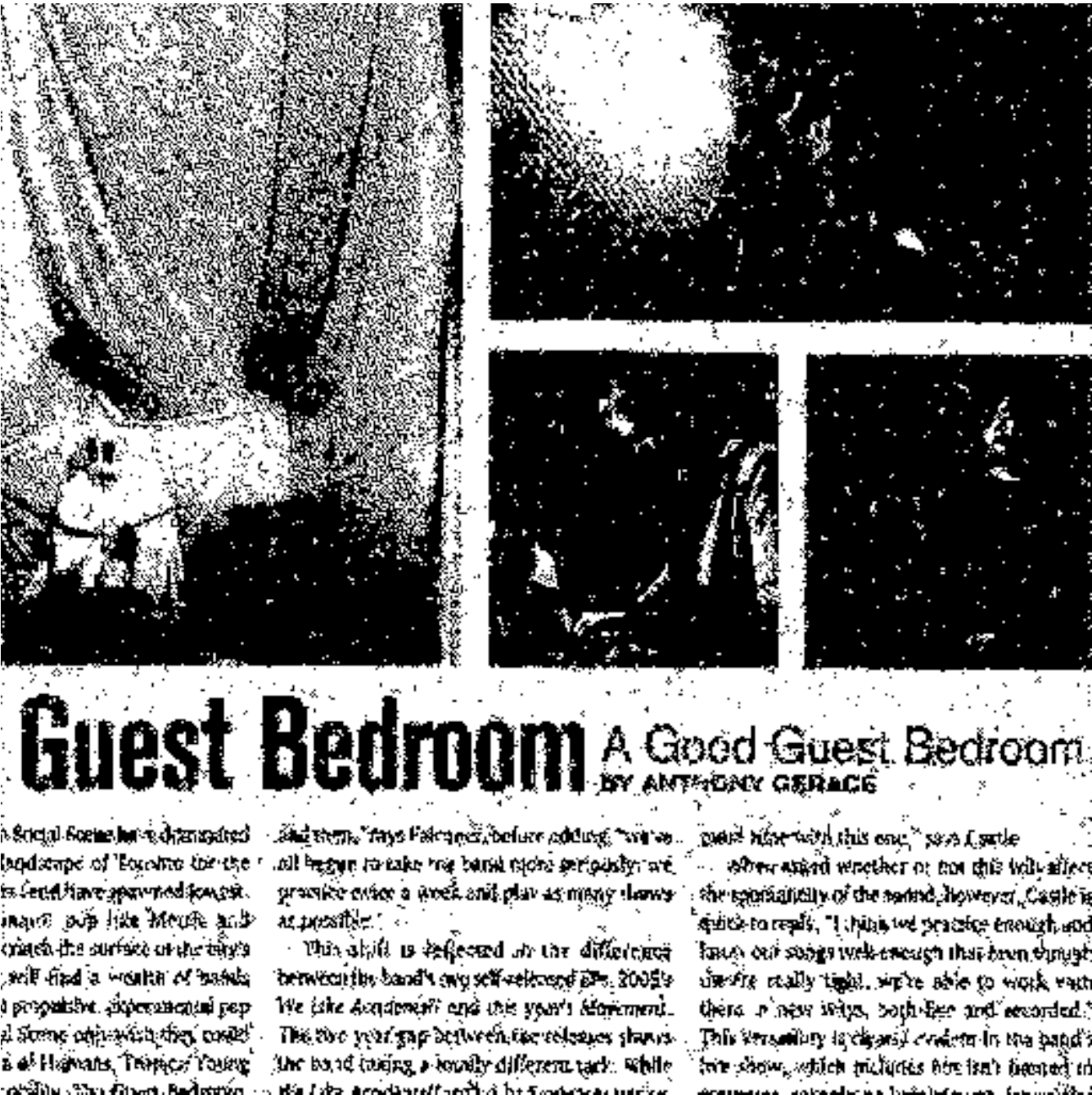}}
\hspace{.2in} \subfigure[2D S-DUDE ($k=4, m=4$)]{\label{magazine 2d sdude}
\includegraphics[width=0.3\textwidth]{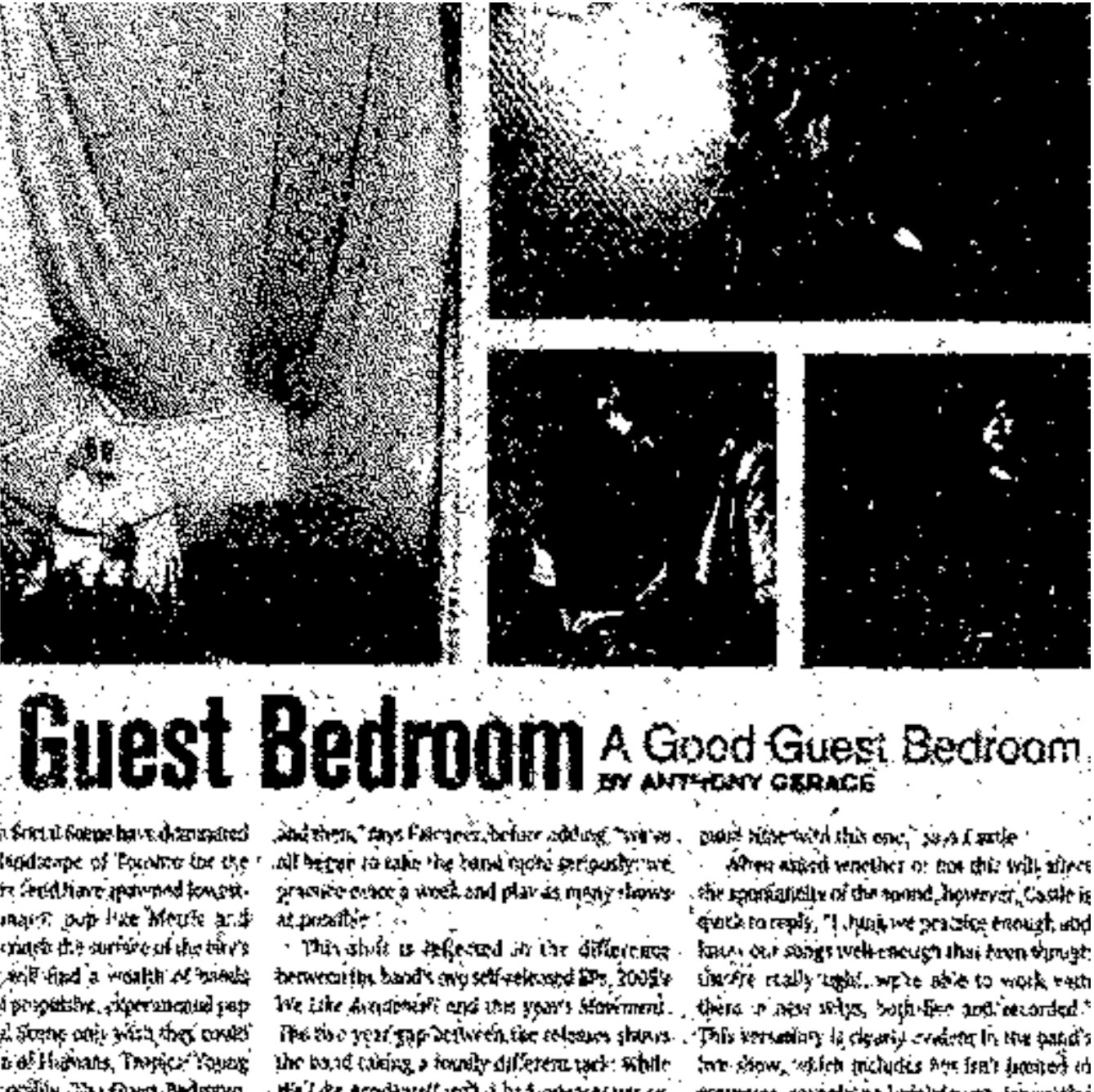}}
\vspace{-.1in}  \caption{Scanned magazine image and denoising results}\label{quadtree}
\end{figure}
Figure \ref{magazine clean} shows the clean binary image obtained from scanning a real magazine page. Unlike the synthetic image in Section \ref{subsec: synthetic}, this image represents the common and realistic characteristics of images that have different textures in different regions of images. The image size is $512\times512$, and  again corrupted by BSC with $\delta=0.1$ that led to the noisy image in Figure \ref{magazine noisy}. 

\begin{figure}[ht]
\centering
\includegraphics[width=0.5\columnwidth]{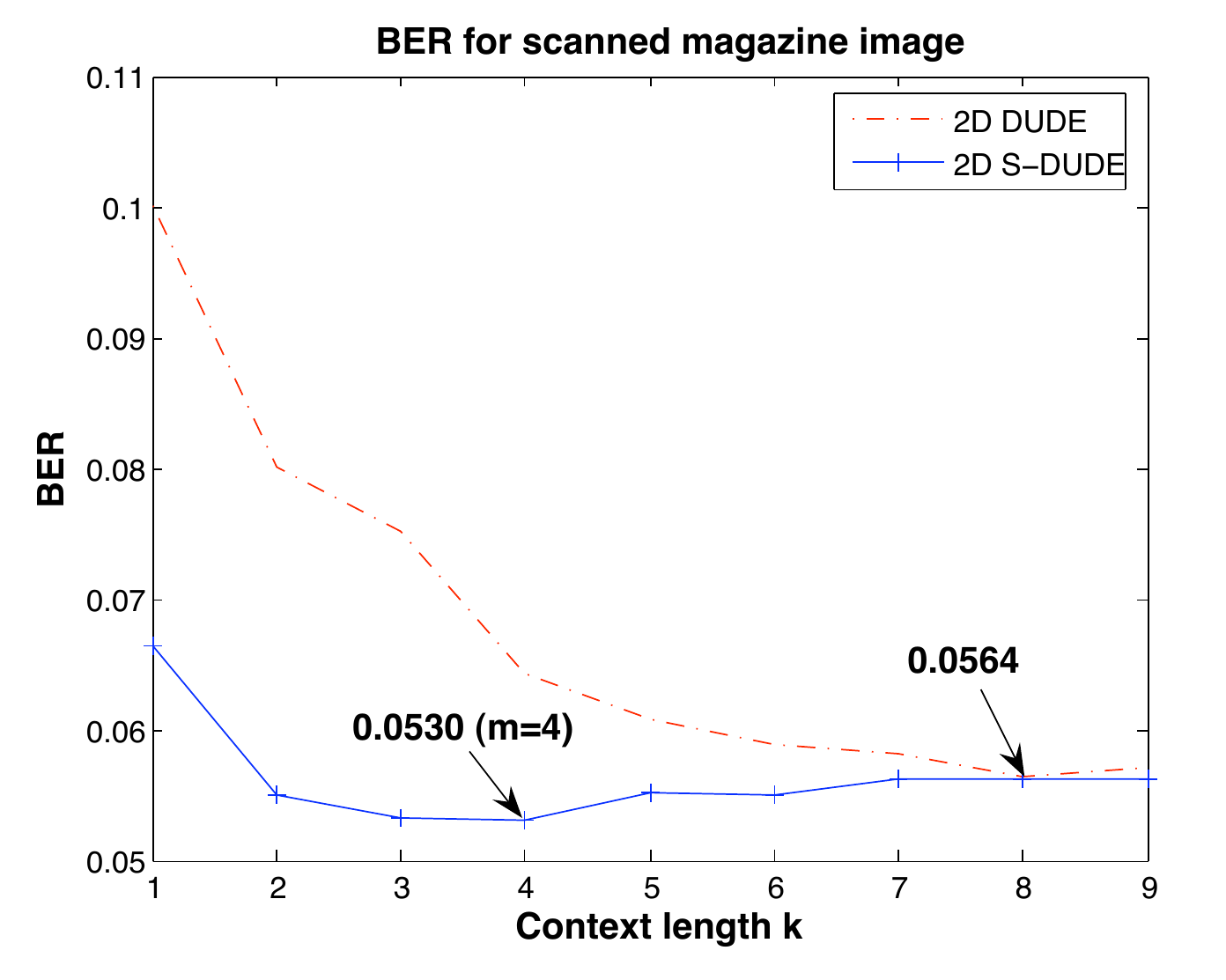}
\caption{Bit error rate comparison for the scanned magazine image}\label{magazine ber}
\end{figure}

Here, we only compare our 2-D S-DUDE with 2-D DUDE,  scheme (i) in the previous subsection, as there were no significant differences between other baselines in Section \ref{subsec: synthetic}, and it is more natural to compare our scheme with one that also uses the 2-D contexts. For our experiments, we varied the context size $k$ from $1$ to $9$ for both of the schemes and tried several values of $m$ for 2-D S-DUDE. The BER plot in Figure \ref{magazine ber} again shows that our 2-D S-DUDE consistently outperforms 2-D DUDE and the best BER is reduced by about 6\%. 
This improvement is significant since 2-D DUDE was already shown to outperform many of the state-of-the-art binary image denoising algorithms. Moreover, we see that the 2-D S-DUDE achieves its optimum performance using context length $k$ which is half of that used by the 2-D DUDE, resulting in overall lower complexity. That is, although 2-D S-DUDE introduces another parameter $m$, the complexity is linear in $m$, and it reduces the dependency on $k$ which contributes exponentially to the complexity. Figure \ref{magazine 2d dude} and Figure \ref{magazine 2d sdude} respectively show the denoising results. We  observe that the 2-D S-DUDE not only has a smaller number of errors, but also does a better job than 2-D DUDE in preserving the sub-image textures. 
\subsection{Lena image}
\begin{figure}[h]
\centering \subfigure[Clean image]{\label{lena clean}
\includegraphics[width=0.25\textwidth]{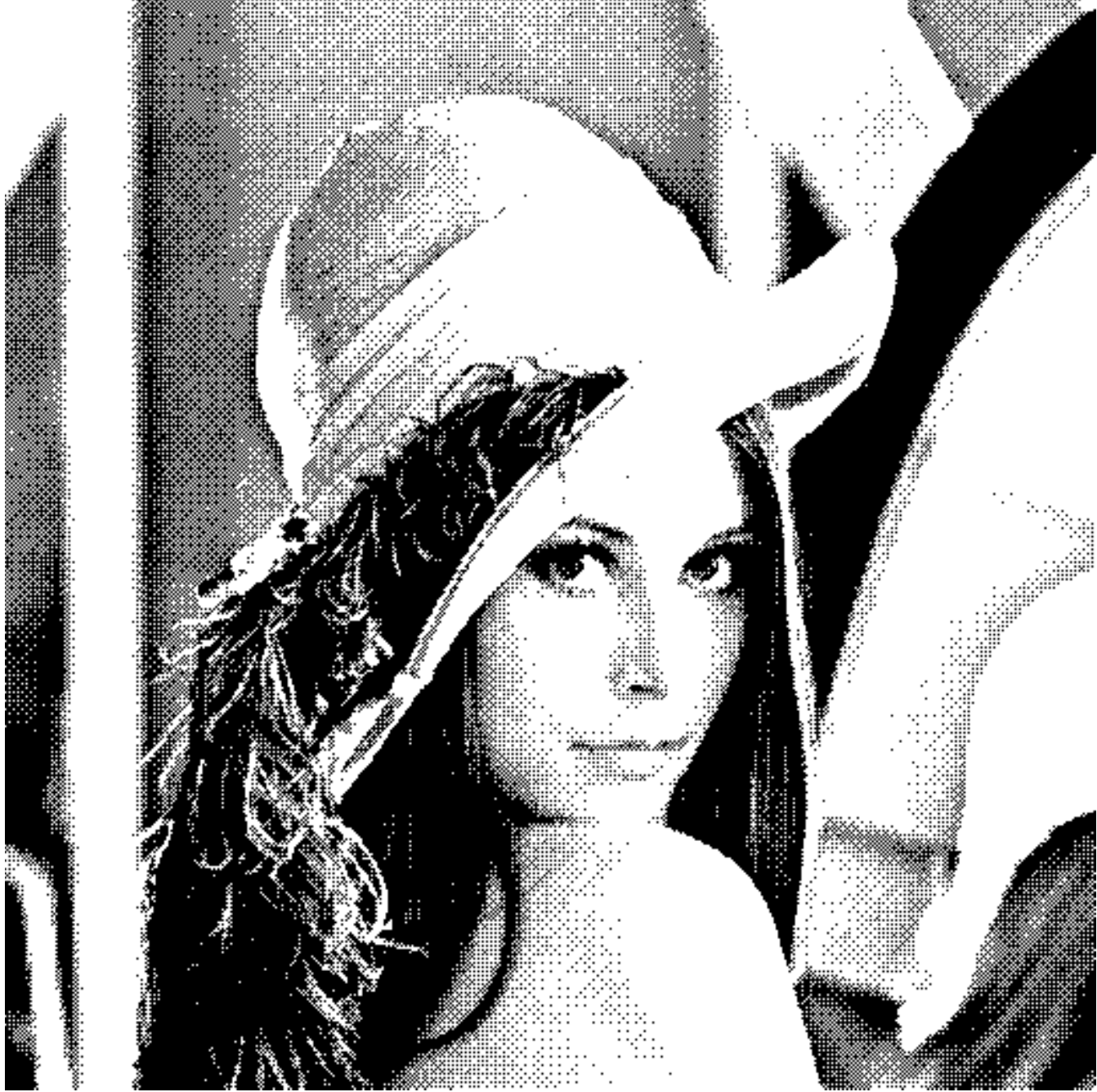}}
\hspace{.1in} \subfigure[Noisy image]{\label{lena noisy}
\includegraphics[width=0.25\textwidth]{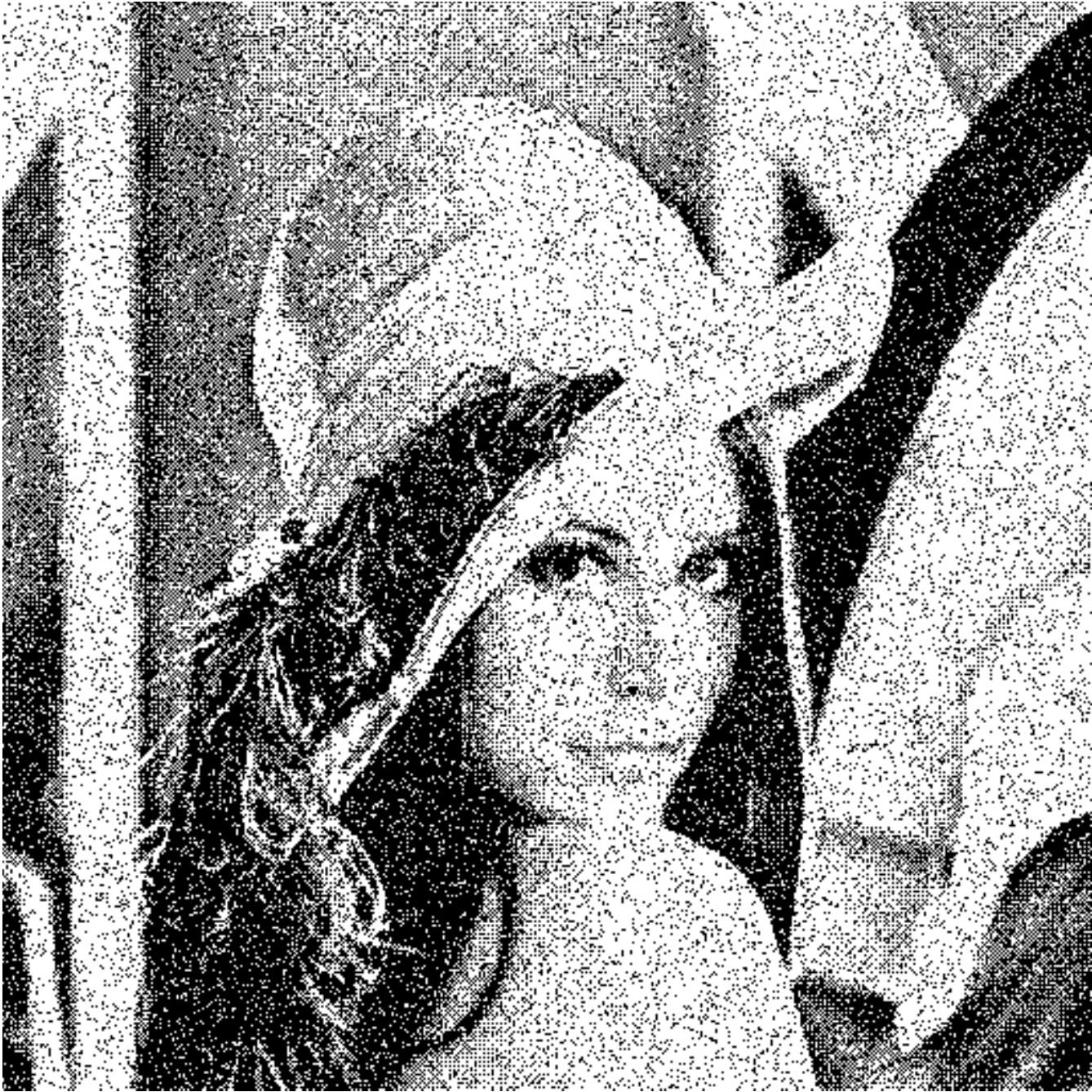}}
\hspace{.1in}\subfigure[2D DUDE($k=6$) and 2D S-DUDE ($k=6, m=4$)]{\label{lena 2d dude}
\includegraphics[width=0.25\textwidth]{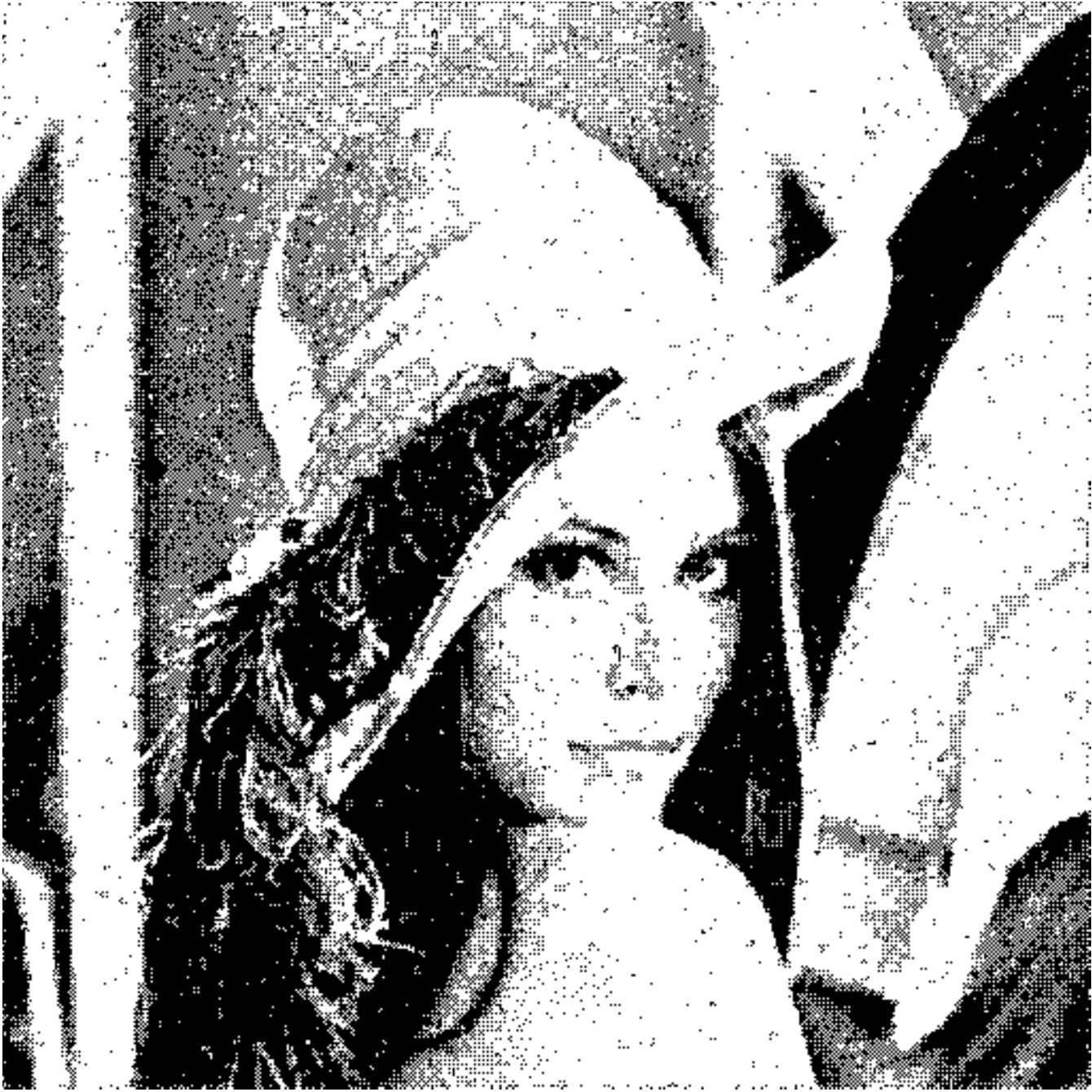}}
\subfigure[BER plot]{\label{lena ber}
\includegraphics[width=0.5\textwidth]{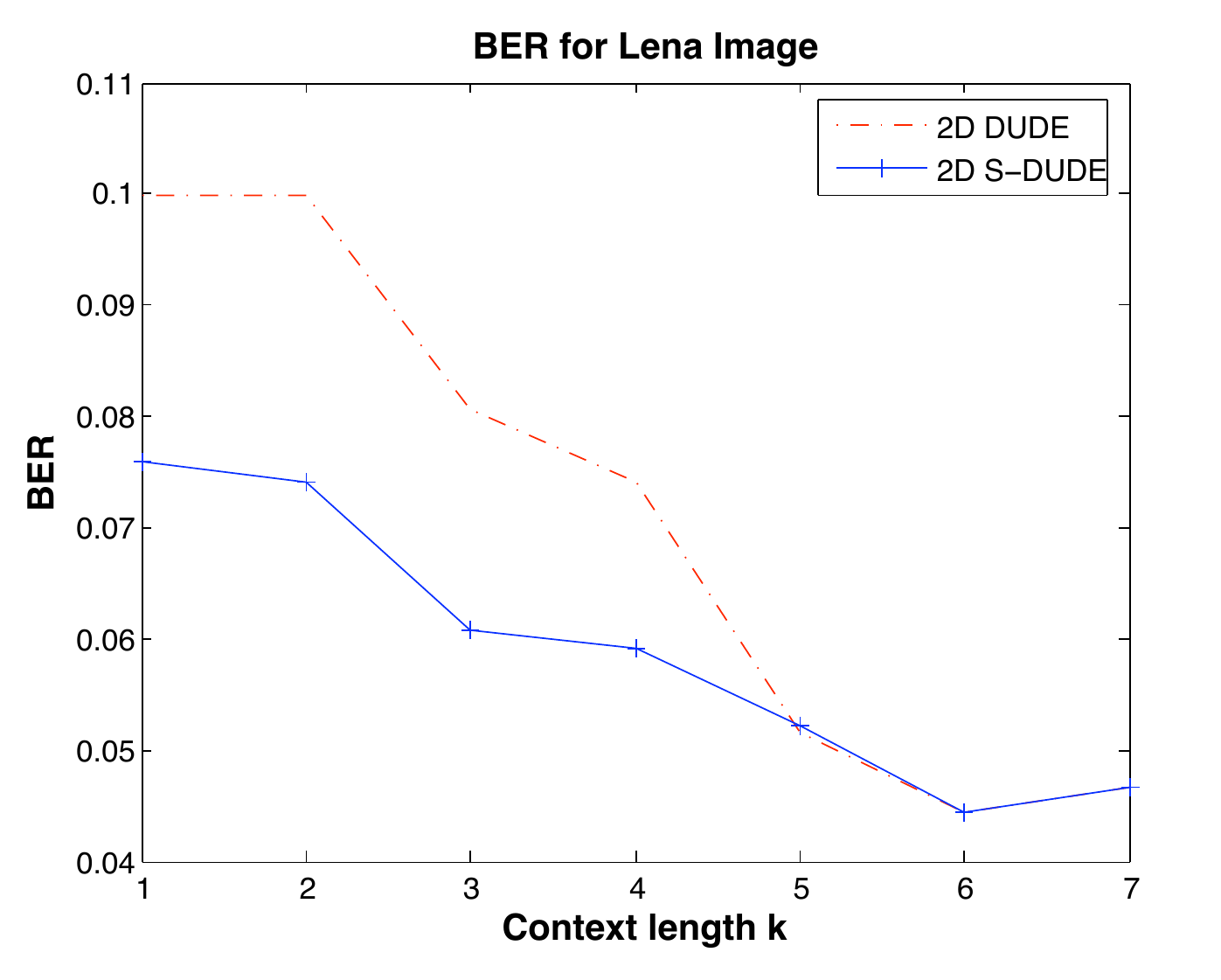}}
\vspace{-.1in}  \caption{Lena image and denoising results}\label{quadtree}
\end{figure}

We now show the results for the  binary Lena image of size $512\times512$. The noisy channel is identical to the previous subsections. Figure \ref{lena clean} and Figure \ref{lena noisy} show the clean and noisy Lena images, and Figure \ref{lena 2d dude} shows the denoising results for both 2-D DUDE with $k=6$ and 2-D S-DUDE with $k=6$ and $m=4$. Figure \ref{lena ber} shows the BER plot for both 2-D DUDE and 2-D S-DUDE. 

These results show what might be expected; when the image characteristics  are largely homogeneous, the best denoising performance of 2-D DUDE and 2-D S-DUDE are similar. However, as we can see in the BER plot in Figure \ref{lena ber}, 2-D S-DUDE reduces the BER faster than 2-D DUDE even with smaller context size $k$ by introducing another parameter $m$. This can be beneficial when we do not know which $k$ would be the optimal for denoising in a practical scenario.

%\begin{figure}[h]
%\centering 
%\includegraphics[width=0.5\textwidth]{Lena_ber_comp}
%\end{figure}

\section{Concluding Remarks}\label{sec: conclusion}
We have generalized the S-DUDE proposed in \cite{sdude} to two-dimensional data. Due to the hardness of optimally segmenting the 2-D data, we introduced a QT decomposition-based reference class of shifting 2-D sliding window denoisers, then utilized the PH scanning technique to efficiently implement the scheme that can attain the optimum performance in the reference class without knowing anything a priori about the characteristics of the underlying clean data. Experimental results show that our scheme can be effective in further reducing the loss of 2-D DUDE, especially for heterogenous images consisting of sub-images of varying natures. Among other related lines of inquiry, future work will investigate the effectiveness of combining more general data segmentation and scanning techniques.

\section*{Acknowledgement}
The authors are grateful to Erik Ordentlich for helpful discussions.

\end{document}